\documentclass[aip,jcp,amsmath,amssymb,floatfix,letterpaper,groupedaddress,
reprint,  
]{revtex4-2}
\usepackage[utf8]{inputenc}         %
\usepackage{graphicx}               
\graphicspath{{.}}
\usepackage[caption=false]{subfig}  
\usepackage{enumitem}               
\usepackage{nicefrac}               
\usepackage{physics}                
\usepackage{siunitx}                
\usepackage{dsfont}                 
\usepackage{booktabs}               
\usepackage[version=4]{mhchem}      
\usepackage{bm}                     
\usepackage{mathtools}              
\usepackage{hyperref}               
\usepackage{lipsum}                 
\usepackage{mwe}                    

\usepackage{color,soul}  

\begin{document}

\title{Time dependent Vibrational Electronic Coupled Cluster (VECC) theory for non-adiabatic nuclear dynamics}

\author{Songhao Bao}
 \email{bsonghao@uwaterloo.ca}  
\author{Neil Raymond}
\author{Marcel Nooijen}
\email{nooijen@uwaterloo.ca}
\affiliation{Department of Chemistry, University of Waterloo, Waterloo, Onario, Canada, N2L 3G1}
\date{\today}

\begin{abstract}
A time-dependent vibrational electronic coupled-cluster (VECC) approach is proposed to simulate photo-electron/UV-VIS absorption spectra, as well as time-dependent properties for non-adiabatic vibronic models, going beyond the Born-Oppenheimer approximation.
A detailed derivation of the equations of motion and a motivation of the ansatz are presented.
The VECC method employs second-quantized bosonic construction operators and a mixed linear and exponential ansatz to form a compact representation of the time-dependent wave-function.
Importantly, the method does not require a basis set, has only few user-defined inputs, and has a classical (polynomial) scaling with respect to the number of degrees of freedom (of the vibronic model), resulting in a favourable computational cost.
In benchmark applications to small models and molecules the VECC method provides accurate results, compared to Multi-Configurational Time-dependent Hartree (MCTDH) calculations when predicting short-time dynamical properties (i.e. photo-elecron / UV-VIS absorption spectra) for non-adiabatic vibronic models. To illustrate the capabilities the VECC method is also applied successfully to a large vibronic model for hexahelicene with 14 electronic states and 63 normal modes, developed in the group by Santoro. ~\cite{aranda2021vibronic} 

\end{abstract}


\maketitle

\section{\label{sec:One}Introduction}

The description of non-adiabatic nuclear dynamical processes is a very active area of investigation.\@
It is at the heart of photochemical processes such as the investigation of intersystem crossings,~\cite{%
bernardi1996potential, ismail2002ultrafast, robb2018theoretical,tully2012perspective,schuurman2018dynamics%
}
like those found in the process of singlet fission.~\cite{%
singh1965laser,swenberg1968bimolecular,geacintov1969effect,merrifield1969fission,merrifield1968theory,johnson1970effects,suna1970kinematics,merrifield1971magnetic,klein1975formation,swenberg1973exciton,zeng2014low, zeng2014seeking, zeng2016design, zeng2016through, pollice2021organic%
}
%
%
In such molecular processes there can be large changes in nuclear geometries and the coupling to the electronic ground state is often important.\@
The most widely applicable approaches in this context employ on-the-fly dynamics (invoking electronic structure calculations of energy gradients and non-adiabatic couplings as needed).\@
Various on-the-fly schemes can be used to approximate the nuclear dynamics on multiple electronic states, for example: Ehrenfest dynamics,~\cite{li2005ab,makhov2017ab, jenkins2018ehrenfest} surface hopping,~\cite{tully1990molecular, richter2011sharc, rauer2016cyclobutane} Ab Initio Multiple Spawning (AIMS),~\cite{ben2000ab, hudock2007ab, williams2021unmasking, mori2012role} and its generalization (GAIMS),~\cite{curchod2016communication} as well as Gaussian wave-packet methods (at various levels of sophistication).~\cite{burghardt2003multiconfigurational, richings2015quantum, christopoulou2021improved,richter2011sharc, richter2012correction, bajo2012mixed, mai2014perturbational, mai2015general, mai2018general}



\begin{figure}\centering%
    \includegraphics[width=\columnwidth]{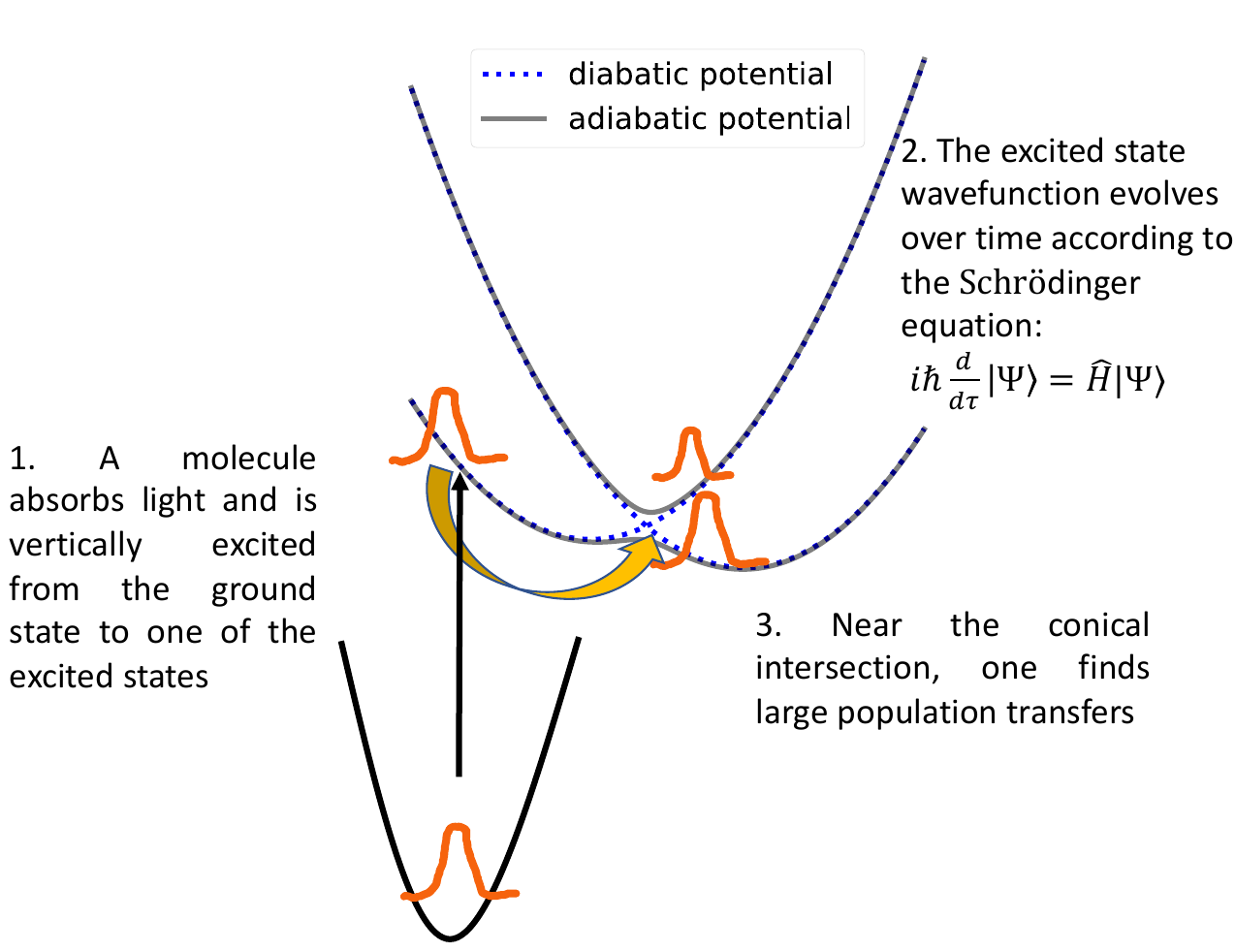}%
    \caption{%
        \label{fig:dynamics_diag}%
        Diagram illustrates how excited state wave-function evolves over time.
    }
\end{figure}

Our primary interest here is the description of electronic spectroscopy including vibrational fine-structure.
The most widely used approaches to include nuclear fine-structure are based on single-surface displaced harmonic oscillator potentials (that include Duschinsky rotations).
Commonly one obtains these potentials through calculations based on the Franck-Condon principle.~\cite{%
grimme2004calculation, hazra2004first, dierksen2005efficient, jankowiak2007vibronic, baiardi2013general, de2019predicting, heldmaier2021uvpd, mashmoushi2021uvpd%
}
However, many systems contain a dense collection of close-lying excited states and the underlying Born-Oppenheimer and harmonic approximations may not be sufficiently accurate to describe the resulting spectra.
In a typical scenario, as shown in \Fref{fig:dynamics_diag}, a molecule absorbs light is vertically excited from the ground state potential energy surface (PES) to an excited state PES, where the excited state wave-function evolves over time.
Near crossing points between these excited PESs (conical intersections), one finds large state population transfer, and the full wave-function becomes a linear combination of the different excited electronic states.
Studying this excited-state dynamics process helps to understand photo-chemical reaction pathways ~\cite{wang2019sudden, etherington2016revealing, zobel2023efficient, shi2019highly}
as well as interpret electronic spectra.~\cite{%
herzberg1966electronic, koppel1984multimode, raab1999molecular, nooijen2003first, hazra2005comparison, hazra2005vibronic, domcke2011conical%
}
For a recent overview see.~\cite{sharma2021vibronically}


A more accurate description of such situations requires the inclusion of multiple coupled electronic states, but the potential energy surfaces need only be accurate over limited spatial extent, close to the ground state equilibrium geometry, as spectroscopic processes occur on short time scales.
One approach for simulating short-time phenomena is the use of a vibronic model in a diabatic basis, where each electronic matrix element can be represented by a short Taylor series expansion (quadratic, or even linear) in the ground state normal mode basis.
Diabatic vibronic models are very useful for extracting short-time dynamics and spectroscopic information.~\cite{%
koppel1984multimode,neugebauer2004vibronic,neugebauer2005vibronic,nooijen2003first, stanton2007vibronic,schuurman2007vibronic,klein2011quantitative,trofimov2020vibronic,yaghoubi2020ultrafast%
}
Standard electronic structure methods for excited states can ``in-principle'' be used to obtain vibronic models, but in practice require modification to extract the necessary vibronic information.
Some of these standard methods include: time-dependent density functional theory (TD-DFT) methods,~\cite{%
stratmann1998efficient, casida2012progress, moore2013longest}
many body perturbation theory (MBPT) based methods~\cite{%
nakano2002quasi, miyajima2006relativistic, ebisuzaki2007efficient},
equation of motion coupled-cluster (EOMCC) methods and similarity transformed EOMCC approaches.~\cite{%
stanton1993equation, bartlett1994applications, nooijen1997new, nooijen1997similarity_1, nooijen1997similarity_2%
}

A variety of approaches are available to construct vibronic models in the diabatic representation.
One approach is diabatization-by-ansatz, e.g.~\cite{%
koppel1981theory, mahapatra2007effects, stanton2007vibronic, faraji2008towards, williams2020complete, trofimov2022vibronic%
}, requiring only adiabatic energies at selected geometries.
A more reliable option for models with many electronic states are diabatization procedures which aim to optimize the resulting diabatic states based on the overlap of adiabatic states (at close lying geometries), or non-adiabatic couplings.~\cite{%
nooijen2003first, hazra2005vibronic, neugebauer2004vibronic, ichino2009quasidiabatic,  klein2011quantitative, faraji2018calculations, zeng2017diabatization, aleotti2021parameterization, yaghoubi2020ultrafast, aranda2021vibronic, green2021fragment, sharma2021vibronically}

After constructing a diabatic vibronic model, one can proceed with the calculation of the absorption spectrum.~\footnote{based on the vibronic Hamiltonian and transition moments (dipole or more general)}
One approach is using a time-independent Lanczos method.~\cite{%
koppel1984multimode,hazra2005vibronic,stanton2007vibronic,ichino2008vibronic, schuurman2008multimode}
This approach can effectively simulate spectra for small to medium-size systems.~\footnote{because the simple form of the vibronic Hamiltonian allows one to use (primitive) harmonic-oscillator basis sets.}
One can also use wave-packet dynamics.
These methods involve the calculation of the time auto-correlation function (ACF) after initial vertical excitation, using the time-dependent Schr{\"o}dinger equation:
\begin{equation}\label{eqn:QMD_vib}
    i\hbar\dv{}{\tau}\ket{\Psi(\tau)}
    = \hat{H}^{vibr}\ket{\Psi(\tau)}
    .
\end{equation}
The multiconfigurational time-dependent Hartree (MCTDH)~\cite{%
beck2000multiconfiguration,worth2000heidelberg,mctdh:package,mctdh2022manual}
approach (and Gaussian wave-packet variations~\cite{burghardt2003multiconfigurational,richings2015practical})
are among the most successful wave-packet dynamics methods.
In particular, the multi-layer version of MCTDH~\cite{%
wang2003multilayer,manthe2008multilayer,vendrell2011multilayer,meng2013multilayer, wang2018regularizing} can be a highly efficient approach, but it does require some judicious choices from the user, and convergence (of the approach) may be hard to demonstrate convincingly.
While the family of MCTDH  methods are highly efficient, they can still be rather costly due to the high dimensionality of the underlying primitive Hilbert space.
The challenges faced by the family of MCTDH methods motivates us to design an alternative wave-packet dynamics approach.

The idea of this work is to design a second-quantized exponential ansatz that parameterizes the wave-function in a compact manner and to extract the time-correlation function by propagating amplitude equations.
The advantage of this approach is that it does not require the introduction of a vibrational (H.O.) basis and the number of parameters scales with the classical degrees of freedom.

In this paper, we develop a new method named  ``Vibrational Electronic Coupled Cluster (VECC)" to (approximately) solve the wave-packet dynamics problem.
VECC is an application of the more general thermal normal-ordered exponential ansatz (TNOE) approach developed by M. Nooijen and S. Bao.~\cite{nooijen2021normal}
TNOE is a general second-quantized operator method for solving both time-dependent and statistical mechanical problems of bosonic and fermionic many-body systems.
VECC can be regarded as a special form of the TNOE family of methods, in which we parameterize the time-dependent wave-function using a second-quantized exponential ansatz.

Attempts to utilize CC methods for solving the bosonic many-body problem can be dated back to the 1990s.
Prasad developed a time-dependent CC method for single-surface multi-dimensional vibrational problem, in which a polynomial expansion of anharmonic potential is applied.~\cite{prasad1988time}
Mukherjee and co-workers proposed the thermal cumulant approach for the statistical mechanics of the multi-dimensional vibrational problem.~\cite{sanyal1992thermal}
In recent years, there is a resurgence of this subject.
Hirata et.\ al.\ \cite{faucheaux2015higher,faucheaux2018similarity} use a similar operator approach, while, earlier, Christiansen~\cite{christiansen2004vibrational} introduces a basis set in  his vibrational coupled-cluster methods (VCC), both aiming at obtaining accurate vibrational frequencies and fundamental transitions.
However, all these methods mentioned above focus on the single-surface vibrational problem.
In this paper, we extend the CC method to the multi-surface non-adiabatic dynamics vibronic problem.

This paper is organized as follows: In \Sref{sec:Two}, we outline the vibrational electronic coupled-cluster (VECC) theory.
In \Sref{sec:Three} the VECC method is used to simulate photo-electron/UV-VIS absorption spectra for a handful of molecular compounds.
We also share an exploration of calculating diabatic state populations using VECC.
In \Sref{sec:Four} we discuss certain implementation details as well as summarize our overall findings.

\section{\label{sec:Two} Theory}

In this section, we introduce the formal VECC theory.
The primary goals we aim to achieve with the VECC theory are the simulation of photo-electron/UV-VIS absorption spectra and the calculation of time-dependent diabatic state populations for gas phase molecular systems.

\subsection{\label{sec:zero zero} General overview of VECC Theory}

Our starting pointing is a vibronic model Hamiltonian that is represented in diabatic electronic states and with the potential truncated up to quadratic terms.
Model parameters can be obtained using standard electronic structure methods and diabatization protocols.
~\cite{koppel1981theory, mahapatra2007effects, stanton2007vibronic, faraji2008towards, williams2020complete, trofimov2022vibronic,
nooijen2003first, hazra2005vibronic, neugebauer2004vibronic, ichino2009quasidiabatic,  klein2011quantitative, faraji2018calculations, zeng2017diabatization, aleotti2021parameterization, yaghoubi2020ultrafast, aranda2021vibronic, green2021fragment, sharma2021vibronically}
The general vibronic model Hamiltonian can be expressed as a sum over electronic states and bosonic construction operators as follows:
\begin{equation}\label{eqn:vibronic_H}
    \begin{split}
    \hat{H}
    &=  \lsum{ab}\ket{a}\bra{b}
    \Bigg(
        h^{a}_{b}
        + \lsum{i} h^{ai}_{b} \{\cop{i}\}
        + \lsum{i} h^{a}_{bi} \{\aop{i}\}
\\  &+  \lsum{ij} h^{ai}_{bj} \{\cop{i}\aop{j}\}
        +\frac{1}{2}\lsum{ij} h^{aij}_{b} \{\cop{i}\cop{j}\}
        +\frac{1}{2}\lsum{ij} h^{a}_{bij} \{\aop{i}\aop{j}\}
    \Bigg),
    \end{split}
\end{equation}
where labels $a,b$ denote electronic surfaces and labels $i,j$ denote vibrational modes.

In VECC theory, we first apply a bosonic second-quantized coupled-cluster (CC) ansatz to parameterize the time-dependent wave-function.
This CC ansatz is then substituted into the time-dependent Schr\"odinger equation (TDSE) to form coupled-cluster (CC) equations-of-motion (EOM) which are first order differential equations over time.
Time dependent properties can be determined by solving coupled-cluster EOM numerically.

Among the time-dependent properties, of particular interest is the time-autocorrelation function (ACF).
The ACF is defined as the overlap between initial wave-function and the wave-function at time $\tau$.
\begin{equation}\label{eqn:acf_definition}
    \begin{split}
        C(\tau)
        &=  \bra{\psi(t=0)} \ket{\psi(t=\tau)}
    \\  &=  \bra{\psi(t=0)} e^{-i\hat{H}\tau} \ket{\psi(t=0)}.
    \end{split}
\end{equation}
This property is of interest because the vibrationally-resolved electronic spectra of non-adiabatic vibronic models can be generated through Fourier transformation of the ACF (provided that the initial wave-function is designed carefully).
\begin{equation}
    C(\tau) \xRightarrow{\text{Fourier transform}}  \text{Spectra}.
\end{equation}

To calculate the ACF we need to evaluate the quantum propagator $e^{-i\hat{H}\tau}$, whose Hamiltonian $\hat{H}$ (defined in \Eref{eqn:vibronic_H}) has multiple electronic surfaces.
This is problematic because the exponential parameterization for multiple surfaces is non-trivial.
We will first examine simpler cases, such that the result can be generalized to multiple surfaces.

In \Sref{sec:temporary_label} we start with simplified single-surface quadratic vibrational models, which serve as a prototype problem for exploring four different realizations of the VECC ansatz.
The preferred realization is then generalized to full multi-surface vibronic models in \Sref{sec:Two_Two}.
Additional modifications are made to the ansatz to improve the accuracy of the wave-function propagation, as well as simplify the EOM.
Our approach for calculating time-dependent properties using this new VECC method is outlined in \Sref{sec:Two_Three}.
We focus on the time-autocorrelation function (ACF) and the diabatic state population, although in principle any time-dependent properties can be obtained.



\subsection{\label{sec:temporary_label} Exploring four single-surface approaches.}

Consider a molecule sited at the minima of its ground state (GS) potential energy surface (PES).
Its dimensionless normal modes are based on the GS Hessian.
The excited-state surface (near the vertical GS geometry) is parameterized as a multidimensional displaced harmonic oscillator (DHO) that includes Duschinsky rotations.
The Hamiltonian that describes this multi-dimension DHO can be expressed as:
\begin{equation}\label{eqn:single_suface_H_in_pq}
    \hat{H} = E_x
    + \lsum{i} \frac{1}{2} \hbar\omega_{i} (\hp{i}{2} + \hq{i}{2})
    + \lsum{ij} V_{ij} \hq{i}\hq{j}
    + \lsum{i} \alpha_{i} \hq{i}
    ,
\end{equation}
where $E_x$, $\omega_{i}$, $\alpha_{i}$, $V_{ij}$ are the vertical energy, frequencies, linear displacements, and quadratic adjustment parameters respectively.
The dimensionless momentum and coordinate operators $\hat{p}_i$ and $\hat{q}_i$ can be expressed in second-quantized form:
\begin{equation}\label{eqn:q_to_second}
    \hq{i} = \frac{1}{\sqrt{2}}(\cop{i}+\aop{i}),
\end{equation}
\begin{equation}\label{eqn:p_to_second}
    i\hp{i} = \frac{1}{\sqrt{2}}(\cop{i}-\aop{i}).
\end{equation}

Substituting \Erefpl{eqn:q_to_second}{eqn:p_to_second} into \Eref{eqn:single_suface_H_in_pq} we can express the multi-dimensional DHO in a second-quantized form:
\begin{equation}\label{eqn:single_surface_H_in_SQ}
    \begin{split}
    \hat{H}
    &=  \left(E_x + \frac{1}{2}\lsum{ii}V_{ii} + \frac{1}{2}\lsum{i}\hbar\omega_{i}\right)
        + \lsum{i}\frac{\alpha_{i}}{\sqrt{2}} \{\aop{i}\}
\\  &+  \lsum{i}\frac{\alpha_{i}}{\sqrt{2}} \{\cop{i}\}
        + \lsum{ij}\left(\hbar\omega_{i}\delta_{ij} + V_{ij}\right) \{\cop{i}\aop{j}\}
\\  &+  \lsum{ij}\frac{V_{ij}}{2} \{\cop{i}\cop{j}\}
        + \lsum{ij}\frac{V_{ij}}{2} \{\aop{i}\aop{j}\}
    ,
    \end{split}
\end{equation}
where the notation ``$\{\}$'' denotes normal ordering with respect to a vacuum state that satisfies $\aop{i}\ket{0} = 0$.~\footnote{%
The initial state in the simulation of absorption spectra is represented by $\ket{0} \mu$, where $\mu$ is a constant representing the electronic transition moment.
For this reason we can focus on an initial state represented by $\ket{0}$.
}

Using the definition of the Hamiltonian in \Eref{eqn:single_surface_H_in_SQ}, we will explore four different single-surface approaches for solving the TDSE.
The guiding principle when designing these approaches is to strike a balance between accuracy and robustness.
To achieve this we compare simulated ACF and absorption spectra of each approach against exact diagonalization (ED) for a small model system.

\subsubsection{\label{sec:Two_One_One}
    Single-exponential approach \texorpdfstring{($\ket{\psi} = \eT \ket{0}$)}{}
}

The first and most apparent ansatz is the single-exponential ansatz:
\begin{equation}
    \ket{\psi} = \eT \ket{0},
\end{equation}
where $\hat{T}$ is truncated at singles and doubles (SD):
\begin{equation}\label{eqn:T_op_def_single_surface}
    \hat{T} = t_0 + \lsum{i}t^{i}\cop{i} + \frac{1}{2}\lsum{ij}t^{ij}\cop{i}\cop{j}.
\end{equation}
Substituting this ansatz into the TDSE, closed-form CC EOM can be formulated
\begin{equation}\label{eqn:single_exp_S1_op_1}
    i\dv{}{\tau}\eT \ket{0} = \hat{H}\eT \ket{0}.
\end{equation}
Multiplying by $\emT$ on both sides of \Eref{eqn:single_exp_S1_op_1} gives us
\begin{equation}\label{eqn:single_exp_S1_op_2}
\begin{split}
    i\dv{\hat{T}}{\tau} \ket{0}
    &= \left( \emT\hat{H}\eT \right)\ket{0}
\\  &= \left( \hat{H}\eT \right)_{\conntext}\ket{0},
\end{split}
\end{equation}
where the subscript ``conn.'' denotes connected.
It can be proved that all disconnected contributions explicitly cancel out in $\emT \hat{H}\eT $ by writing the string of operators in normal ordering and applying the Baker-Campbell-Hausdorf (BCH) series expansion.
This is in analogy to the usage of single-reference Coupled Cluster theory in electronic structure.~\cite{crawford2007introduction}

Next, we introduce a projection manifold $\bra{0} \proj $
\begin{equation}\label{eqn:proj_manifold_def}
    \proj = \{\identity, \aop{i}, \aop{i}\aop{j} \},
\end{equation}
and project on both sides of \Eref{eqn:single_exp_S1_op_2}
\begin{equation}\label{eqn:single_exp_S1_proj}
    i\bra{0} \proj \dv{\hat{T}}{\tau} \ket{0}
    = \bra{0} \proj \left(\hat{H}\eT \right)_{\conntext} \ket{0} .
\end{equation}
Coupled-cluster amplitudes can then be determined by numerically solving the set of ODEs in \Eref{eqn:single_exp_S1_proj}.
The time-autocorrelation function in these single surface cases is obtained from the projection against $\bra{0}$, or the $\proj = \identity$ component.

\par
\begin{figure}[!h]\centering%
    \subfloat[\label{subfig:one_a}%
        Time-autocorrelation functions of ED and single-exponential ansatz.%
    ]{
        \includegraphics[width=0.8\columnwidth]{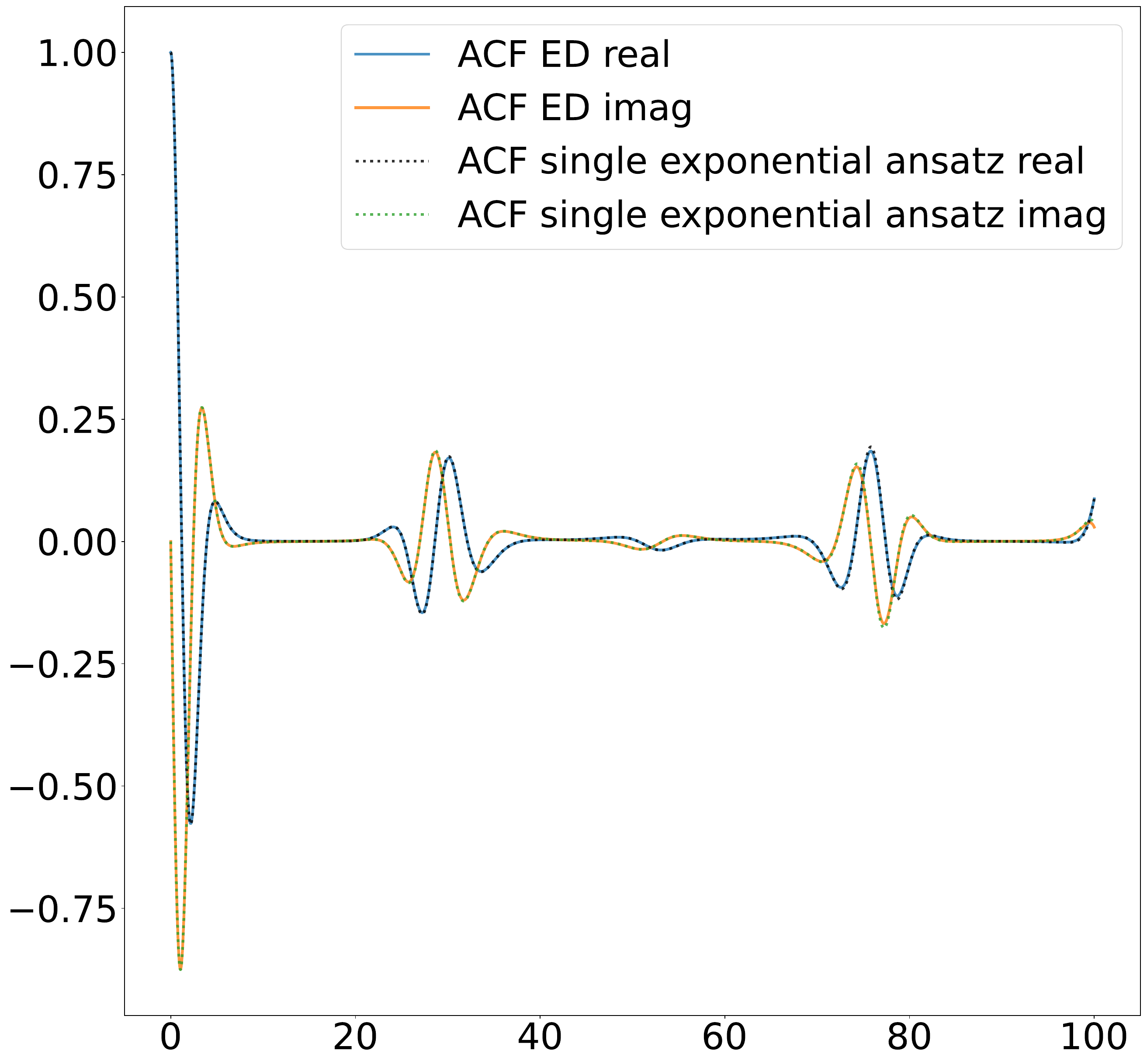}%
    }\hfill
    \subfloat[\label{subfig:one_b}%
        Absorption spectras of ED single-exponential ansatz.%
    ]{%
        \includegraphics[width=0.8\columnwidth]{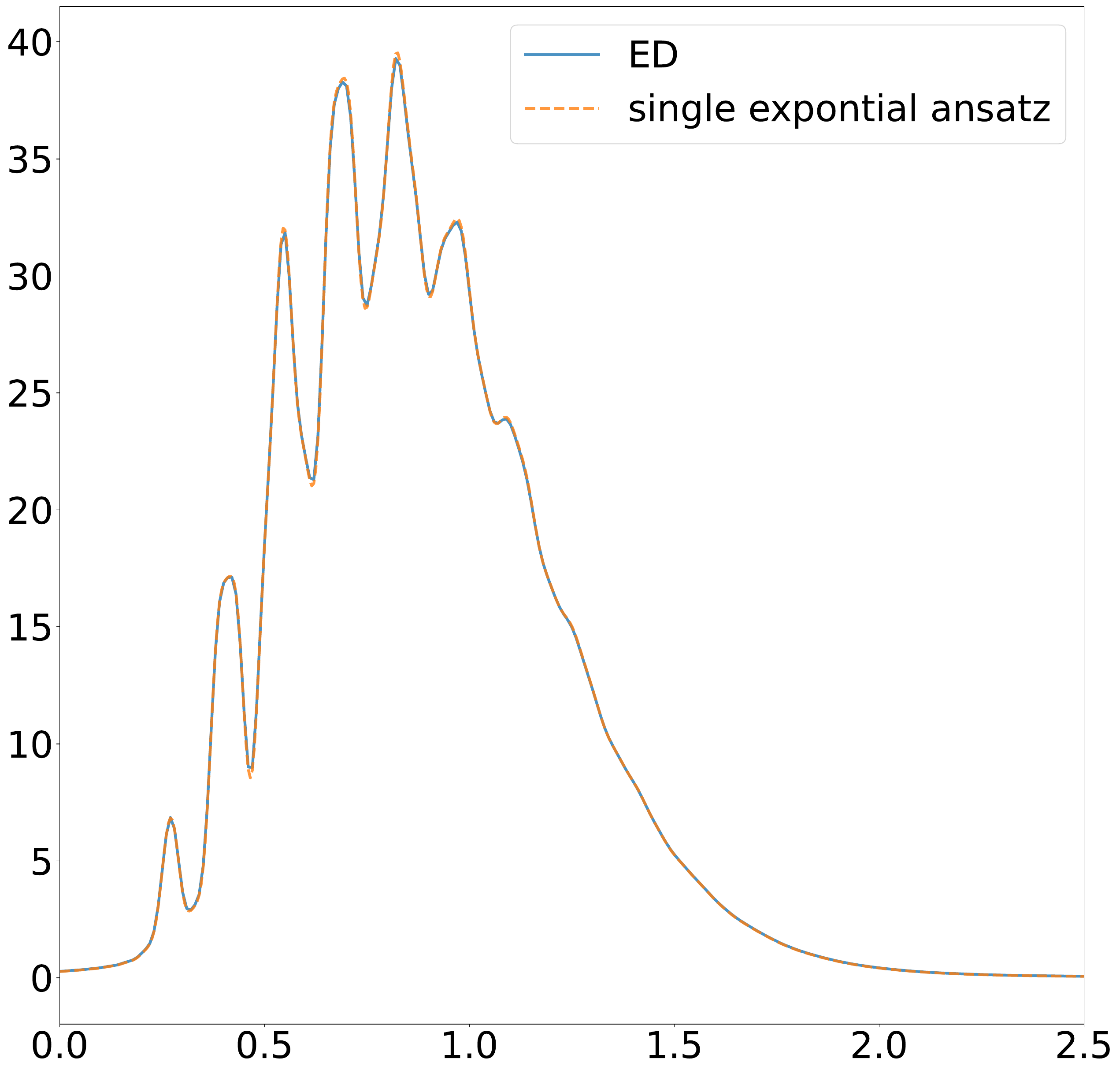}
    }\hfill
    \caption{%
        \label{fig:single_surface_model}%
        For a single-surface vibrational model the single-exponential ansatz produces a numerically exact time-autocorrelation function and its Fourier transformed spectra when compared to ED, and a SD truncation.
    }

\end{figure}

To evaluate the single-exponential approach we will compare to exact diagonalization (ED), which is a numerically converged solution in a (large) finite basis.
First we calculate an ACF.
We then use the \texttt{autospec84} module in the MCTDH suite to Fourier transform the ACF resulting in a simulated absorption spectra.~\cite{beck2000multiconfiguration}
These results are presented in~\Fref{fig:single_surface_model}.
For the single-exponential approach, the simulated ACF and absorption spectra agree extremely well with the ED result.

The appealing feature of this single-exponential ansatz approach, is that it is numerically exact for these quadratic model Hamiltonians.
The formal proof of this conclusion is that the CC residual equations are in closed form and the CC amplitudes truncate explicitly at singles and doubles.
As shown in \Eref{eqn:Proof_close}, when applying Wick's theorem to evaluate triple residual, there are no surviving contractions, provided that $\hat{H}$ is at most quadratic.
\begin{equation}
\label{eqn:Proof_close}
    \bra{0} \aop{i}\aop{j}\aop{k} \left(\hat{H}\eT \right)_{\conntext} \ket{0} = 0
\end{equation}

Although with this appealing feature of numerical exact for the single surface quadratic Hamiltonian, we find difficulties to generalize the single-exponential ansatz approach to the full vibronic coupling problem due to the non-unitary property of it.
In the single-exponential ansatz, we essentially parameterize the quantum propagator $e^{-i\hat{H}\tau}$, as it acts on the reference state
\begin{equation}
    \eT \ket{0} \equiv e^{-i\hat{H}\tau} \ket{0}
\end{equation}
The quantum propagator $e^{-i\hat{H}\tau}$ is an unitary operator while the exponential ansatz is non-unitary.
For single-surface harmonic vibrational problems the norm of time-dependent wave-function and the energy expectation value are conserved, as they are numerically exact.
The constant term $t_0$ in the operator $\hat{T}$ (\Eref{eqn:T_op_def_single_surface}) plays a vital role in this regard.




However, we found that the non-unitary property of the single-exponential ansatz is potentially problematic when extending to full vibronic models.
Due to non-vanishing disconnected contributions in the CC residual equations the resulting CC EOM are no longer closed, and therefore the single-exponential ansatz (with singles and doubles truncation) becomes an approximate method rather than numerically exact.
When applying the single-exponential ansatz to the full vibronic models, we observed divergence of the norm and energy expectation values.~\footnote{Divergence was not guaranteed, but with enough regularity (> 90\% of our testing) divergence was observed after propagating the ACF longer than 50fs (in the full vibronic models)}
As a consequence, both the simulated CC amplitude and the ACF diverges from our ED result (obtained using a converged finite basis).

To resolve this divergence issue, we propose another ansatz (Linear CI) that has a unitary parameterization and therefore when extending to full vibronic models the norm and energy expectation values are conserved.

\subsubsection{\label{sec:Two_One_Two} Linear CI approach \texorpdfstring{($\ket{\psi}=\hat{Z} \ket{0}$)}{}}

The second ansatz we introduce is the linear configurational interaction (CI) ansatz,
\begin{equation}
    \ket{\psi} = \hat{Z} \ket{0},
\end{equation}
whose CI operator is truncated at singles and doubles (CISD):
\begin{equation}
    \hat{Z} = z_0 + \lsum{i} z^{i} \cop{i} + \frac{1}{2}\lsum{ij} z^{ij} \cop{i}\cop{j}.
\end{equation}
Applying the linear CI ansatz to the TDSE, we form the EOM:
\begin{equation}\label{eqn:linear_CI_EOM}
    i\bra{0} \proj \dv{\hat{Z}}{\tau} \ket{0}
    = \bra{0} \proj \hat{H} \hat{Z} \ket{0}.
\end{equation}

\begin{figure}[!h]\centering%
    \subfloat[\label{subfig:two_a}%
        Time-autocorrelation functions of ED and linear CI ansatz.%
    ]{
        \includegraphics[width=0.8\columnwidth]{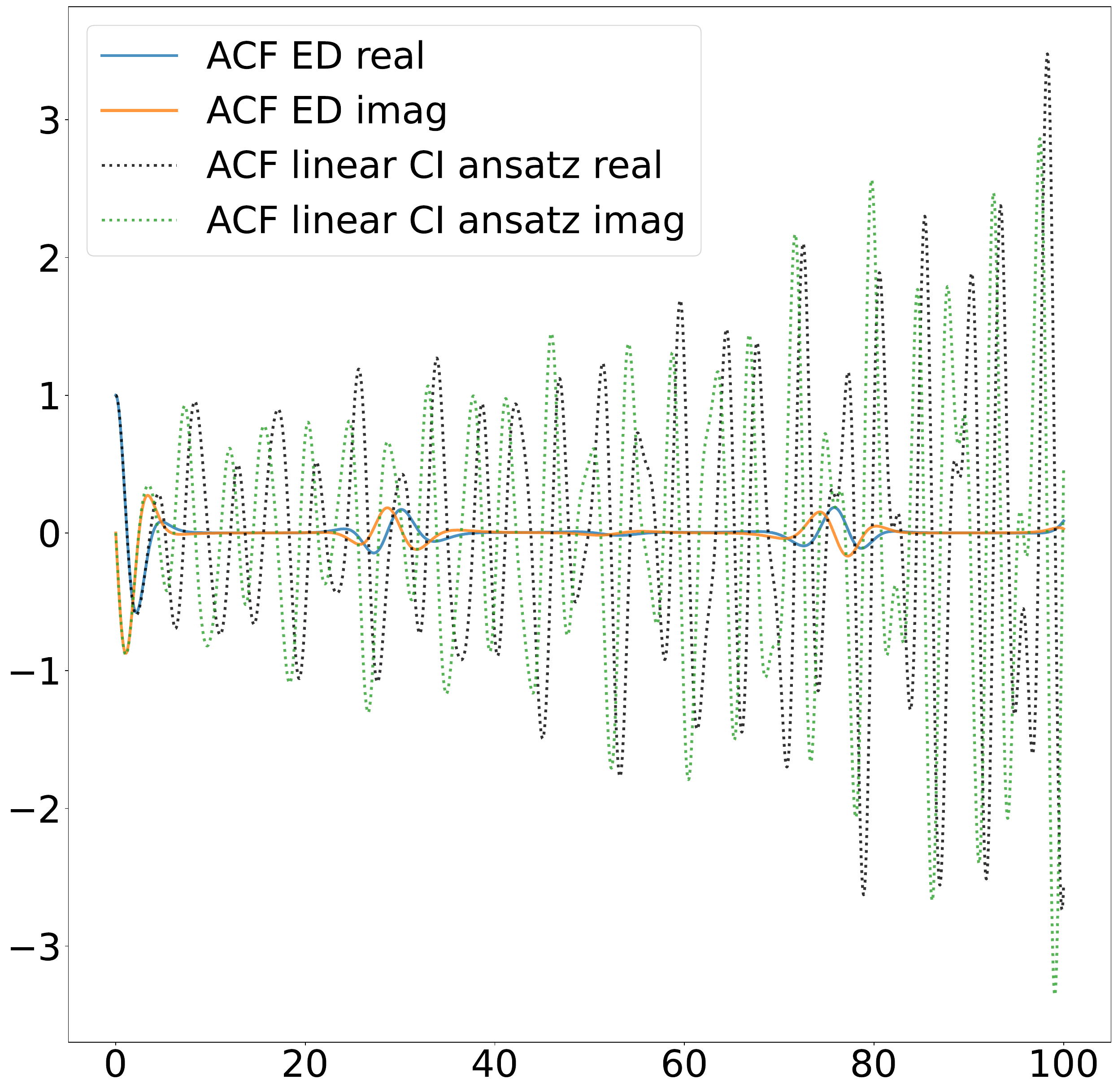}%
    }\hfill
    \subfloat[\label{subfig:two_b}%
        Absorption spectras of ED and linear CI.%
    ]{%
        \includegraphics[width=0.8\columnwidth]{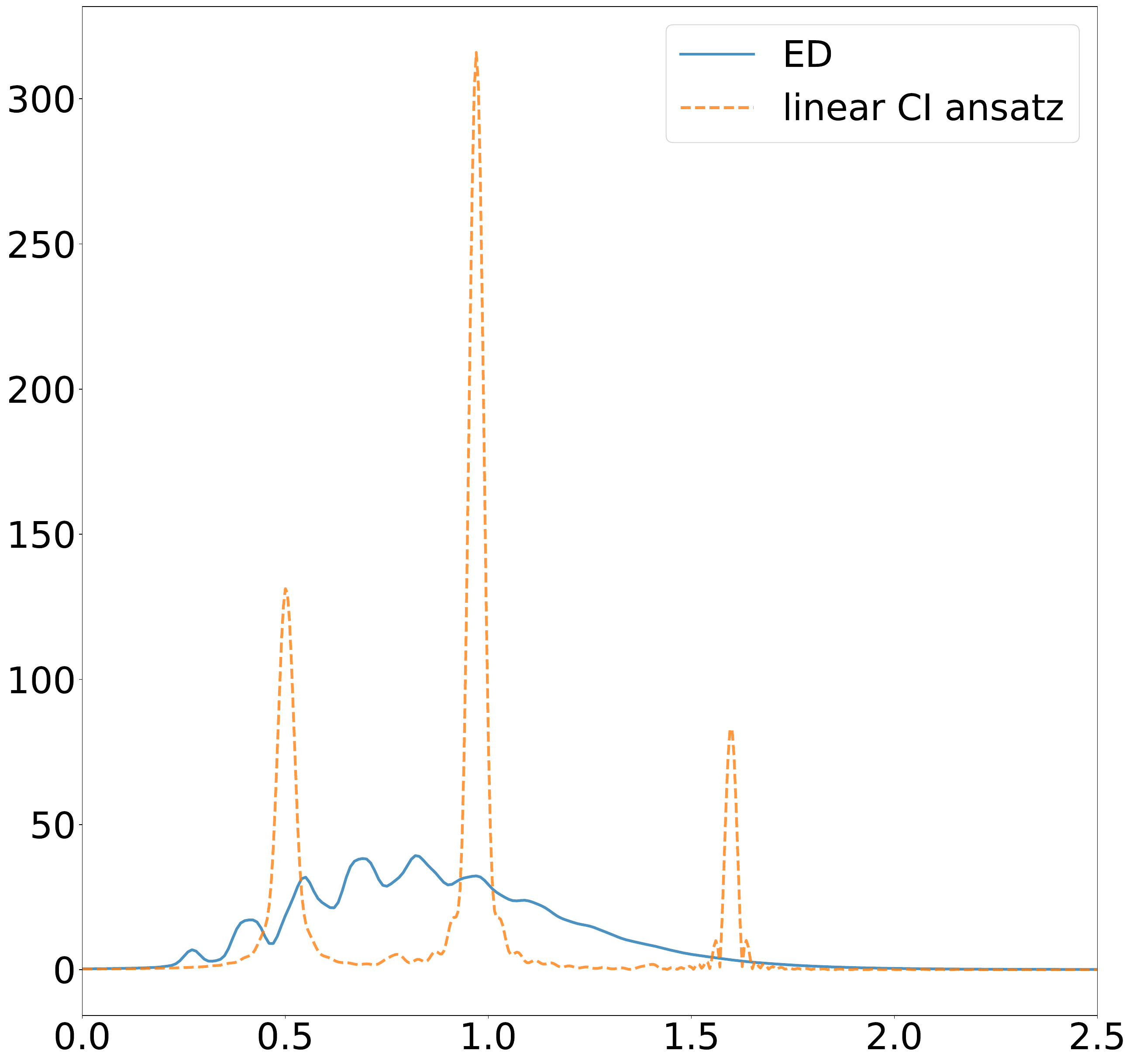}
    }\hfill
    \caption{%
        \label{fig:two}%
        Linear CI ansatz is unitary, but no longer numerical exact with SD truncation.
        Both subfigures show large deviation from ED.
        This is expected since the linear CISD ansatz contains only two fold excitations.
    }
\end{figure}

As shown in \Fref{fig:two}, it turns out that  CISD is a poor approximation to the exact solution of the Schr\"odinger equation.
Both the time-autocorrelation function and the Fourier transformed spectra have large deviations when compared to ED.\@
This result is expected, since the CISD approximation is essentially equivalent to exact diagonalization of the full Hamiltonian in a H.O. basis truncated to only two-fold excitations.
Such a basis set is too small to produce converged results for energy eigenvalues.
Unlike the single-exponential ansatz, which is fully connected and numerically exact, the linear CI ansatz is not explicitly connected and also very approximate.\@
In contrast to CCSD, in CISD,  the triples residuals are non-zero in the singles and doubles approximation.
In \Eref{eqn:linear_CI_EOM}, the right hand side of the equation is not connected and as a result the exact CI equations do not rigorously truncate ever for Bosonic systems.

Although the linear CI ansatz has insufficient accuracy, its unitary property plays an important role when extending to general vibronic models, since it dramatically improves the numerical robustness of time integration.

Recall that our goal is to design an ansatz that keeps a balance between accuracy and robustness.\@
We have seen that the single-exponential ansatz is accurate but not robust (for multi-surface models), while the linear CI ansatz is robust but not accurate.
Therefore, we propose another ansatz that combines the good features of these two ans{\"a}tze, attempting to improve both accuracy and robustness.

\subsubsection{\label{sec:Two_One_Three} Mixed CC/CI approach \texorpdfstring{($\ket{\psi}=\eT \hat{Z}\ket{0}$)}{}}

The mixed CC/CI ansatz is a hybridization of the single-exponential ansatz and the linear CI ansatz.
It is defined as follows
\begin{equation}\label{eqn:_mix_CC_CI_ansatz}
    \ket{\psi} = \eT \hat{Z} \ket{0},
\end{equation}
where
\begin{equation}
    \hat{T} = \lsum{i} t^{i}\cop{i},
\end{equation}
and
\begin{equation}
    \hat{Z} = z_0 + \lsum{i} z^{i}\cop{i} + \frac{1}{2}\lsum{ij} z^{ij}\cop{i}\cop{j}.
\end{equation}
Substituting this ansatz into the TDSE and projecting with $\bra{0} \proj \emT $ (where $\proj $ is defined in \Eref{eqn:proj_manifold_def}) results in new EOM:
\begin{equation}\label{eqn:mix_CC_CI_EOM_1}
    \bra{0} \proj \left( i\dv{\hat{T}}{\tau}\hat{Z} + i\dv{\hat{Z}}{\tau} \right) \ket{0}
    = \bra{0} \proj \bar{H} \hat{Z} \ket{0},
\end{equation}
where
\begin{equation}\label{eqn:sim_tran_H_single_surface}
    \bar{H} = e^{- \hat{T}} \hat{H} e^{\hat{T}} = \left( \hat{H}\eT \right)_{\conntext} .
\end{equation}

The left hand side of \Eref{eqn:mix_CC_CI_EOM_1} has two unknown terms ($\dv{\hat{T}}{\tau}$ and $\dv{\hat{Z}}{\tau}$) and therefore does not have a unique solution.
Since the manner in which these terms are specified is essentially redundant, there is freedom to choose the manner in which we partition the right hand side, or ``net residual'', of \Eref{eqn:mix_CC_CI_EOM_1} in terms of the $\hat{T}$ and $\hat{Z}$ residuals.
With this freedom, we choose to specify the EOM for $\hat{T}$  first, and then by definition the EOM for $\hat{Z}$  is the remaining part of \Eref{eqn:mix_CC_CI_EOM_1}

We specify the EOM for $\hat{T}$ in the same manner as traditional CC singles EOM (as shown in~\Eref{eqn:single_exp_S1_proj}) and can then form EOM for $\hat{T}$:
\begin{equation}\label{eqn:mix_CC_CI_EOM_2}
    i\dv{t^i}{\tau} = \bar{H}^{i}.
\end{equation}
By substituting \Eref{eqn:mix_CC_CI_EOM_2} into \Eref{eqn:mix_CC_CI_EOM_1}, we can form EOM for $\hat{Z}$:
\begin{equation}\label{eqn:mix_CC_CI_EOM_3}
    i\bra{0} \proj \dv{\hat{Z}}{\tau} \ket{0}
    = \bra{0} \proj \hat{G}\hat{Z} \ket{0},
\end{equation}
where
\begin{equation}
    \hat{G} = \bar{H} - \sum_i \bar{H}^{i}.
\end{equation}

To determine the $\hat{T}$ and $\hat{Z}$ amplitudes as functions of time, we solve the coupled ODEs in \Eref{eqn:mix_CC_CI_EOM_2} and \Eref{eqn:mix_CC_CI_EOM_3} through the numerical real time integration.
Each iteration of the numerical integration is comprised of three stages.
First, we perform the similarity transform of the Hamiltonian in \Eref{eqn:sim_tran_H_single_surface}.
Then, we equate the $\bar{H}^{i}$ element of the similarity transformed Hamiltonian $\bar{H}$ to  $\dv{\hat{T}^i}{\tau}$  based on \Eref{eqn:mix_CC_CI_EOM_2}.
Lastly, we determine $\dv {\hat{Z}}{\tau}$  using \Eref{eqn:mix_CC_CI_EOM_3} and propagate the differential EOM.

\begin{figure}[!h]\centering%
    \subfloat[\label{subfig:three_a}%
        Time-autocorrelation functions of ED and mixed CC/CI ansatz.%
    ]{
        \includegraphics[width=0.8\columnwidth]{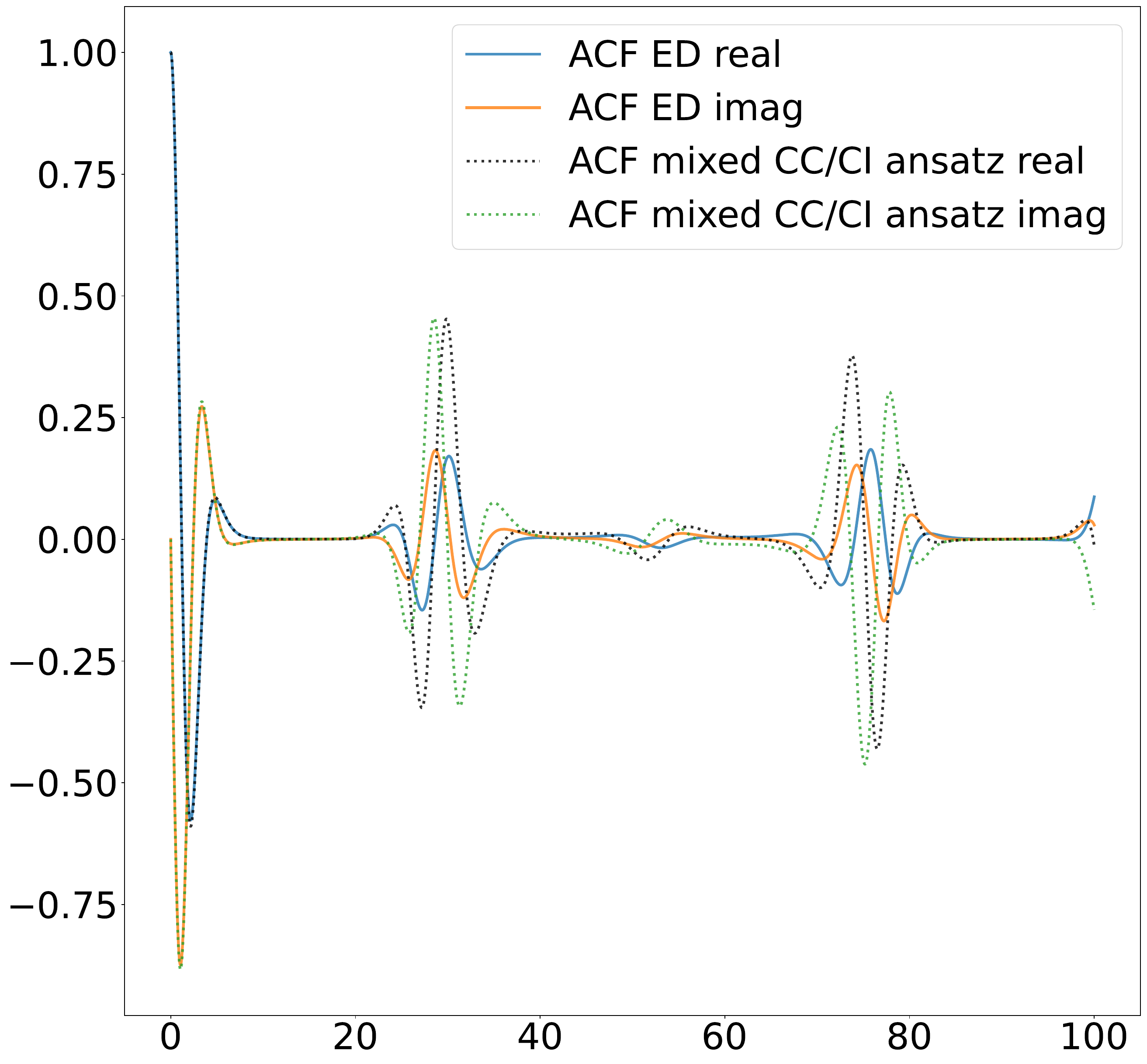}%
    }\hfill
    \subfloat[\label{subfig:three_b}%
        Absorption spectras of ED and mixed CC/CI ansatz.%
    ]{%
        \includegraphics[width=0.8\columnwidth]{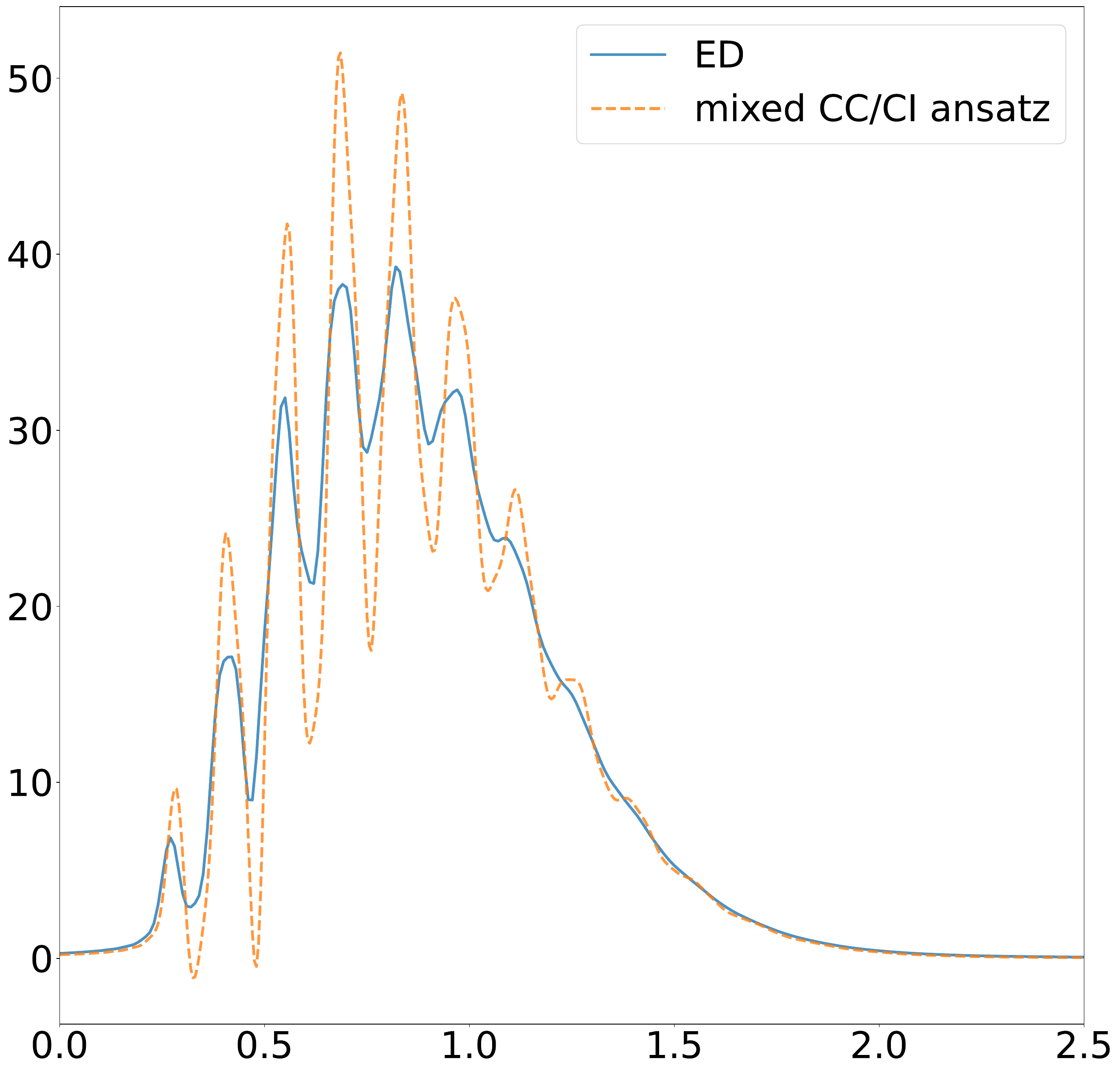}
    }\hfill
    \caption{%
        \label{fig:three}%
        Although mixed CC/CI ansatz is not numerical exact, it agrees fairly closely with ED, showing dramatic improvement compared to the linear CI ansatz, because the linear coupling constants are the dominant terms in the Hamiltonian.
    }
\end{figure}

As shown in \Fref{fig:three}, the mixed CC/CI ansatz is no longer numerically exact.
As in the case of linear CI ansatz, the EOM for $Z$ in \Eref{eqn:mix_CC_CI_EOM_3} are not in a closed form and the $\hat{Z}$ amplitude equation does not strictly truncate at any excitation level.
However, it is not surprising that the mixed CC/CI ansatz can reach considerable accuracy when compared to ED.
This is because the linear coupling constants are the dominant terms in the vibrational model Hamiltonian, and the $\hat{T}$ amplitude, which has an exponential parameterization, accounts for the linear coupling contributions in the CC EOM.
Meanwhile, $\hat{Z}$'s amplitude mainly takes account of remaining quadratic coupling contributions in the time-dependent wave-function, which are much smaller than the linear coupling contributions.

When applied to vibronic models, the mixed CC/CI ansatz improves the robustness when compared to the single-exponential ansatz, because the $\hat{Z}$ component of the ansatz has a linear parameterization which is unitary.

However, it turns out that the mixed CC/CI ansatz can still have numerical divergence issues when being directly extended to full vibronic models.
In the next section, we introduce one final refinement based on mixed CC/CI ansatz that further improves both accuracy and robustness.

\subsubsection{\label{sec:Two_One_Four} Modified projection manifold approach \texorpdfstring{($\bra{0} \proj  \eTd $)}{}}

For the final approach we adopt the mixed CC/CI ansatz described in the previous section, but modify the projection manifold by introducing the operator $\eTd$
\begin{equation}
    \bra{0}\proj \eTd \emT,
\end{equation}
which is applied to both sides of the EOM in \Eref{eqn:mix_CC_CI_EOM_1}
\begin{equation}\label{eqn:mod_proj_EOM_1}
    \bra{0} \proj \eTd
        \left(
            i\dv{\hat{T}}{\tau}\hat{Z}
            + i\dv{\hat{Z}}{\tau}
        \right)
    \ket{0}
    = \bra{0} \proj \eTd \bar{H} \hat{Z} \ket{0}
\end{equation}

We keep the EOM for $\hat{T}$ amplitudes the same as \Eref{eqn:mix_CC_CI_EOM_2} and then substitute them into \Eref{eqn:mod_proj_EOM_1} to form EOM for $\hat{Z}$ amplitudes.
\begin{equation}\label{eqn:proj_EOM_2}
    i\bra{0} \proj \dv{\hat{Z}}{\tau} \ket{0}
    =   \bra{0} \proj \eTd \hat{G} \hat{Z} \ket{0}
        -i\bra{0}
            \proj \left(\eTd - \identity\right)
            \dv{\hat{Z}}{\tau}
        \ket{0}
\end{equation}

The numerical integration procedure, similar to the previous section, is comprised of three stages for each iteration.
A complication, however, is that there are more terms in the $\hat{Z}$ residual equations.
To evaluate the $\dv{\hat{Z}}{\tau}$ term, we need to solve the equation in a descending sequence, from highest order of truncation to lowest order truncation.
This is because the higher order $\dv{\hat{Z}}{\tau}$ operators contribute to the lower rank sector when multiplied by  $e^{\hat{T}^{\dagger}}$, and they need to be consecutively substituted into lower order $\hat{Z}$ residuals.

Comparing to the EOM of $\hat{Z}$ in the previous section \Eref{eqn:mix_CC_CI_EOM_3}, the introduction of the modified projection manifold complicates the EOM of $\hat{Z}$.

\begin{figure}[!h]\centering%
    \subfloat[\label{subfig:four_a}%
        Time-autocorrelation functions of ED and mixed CC/CI ansatz with modified projection manifold.%
    ]{
        \includegraphics[width=0.8\columnwidth]{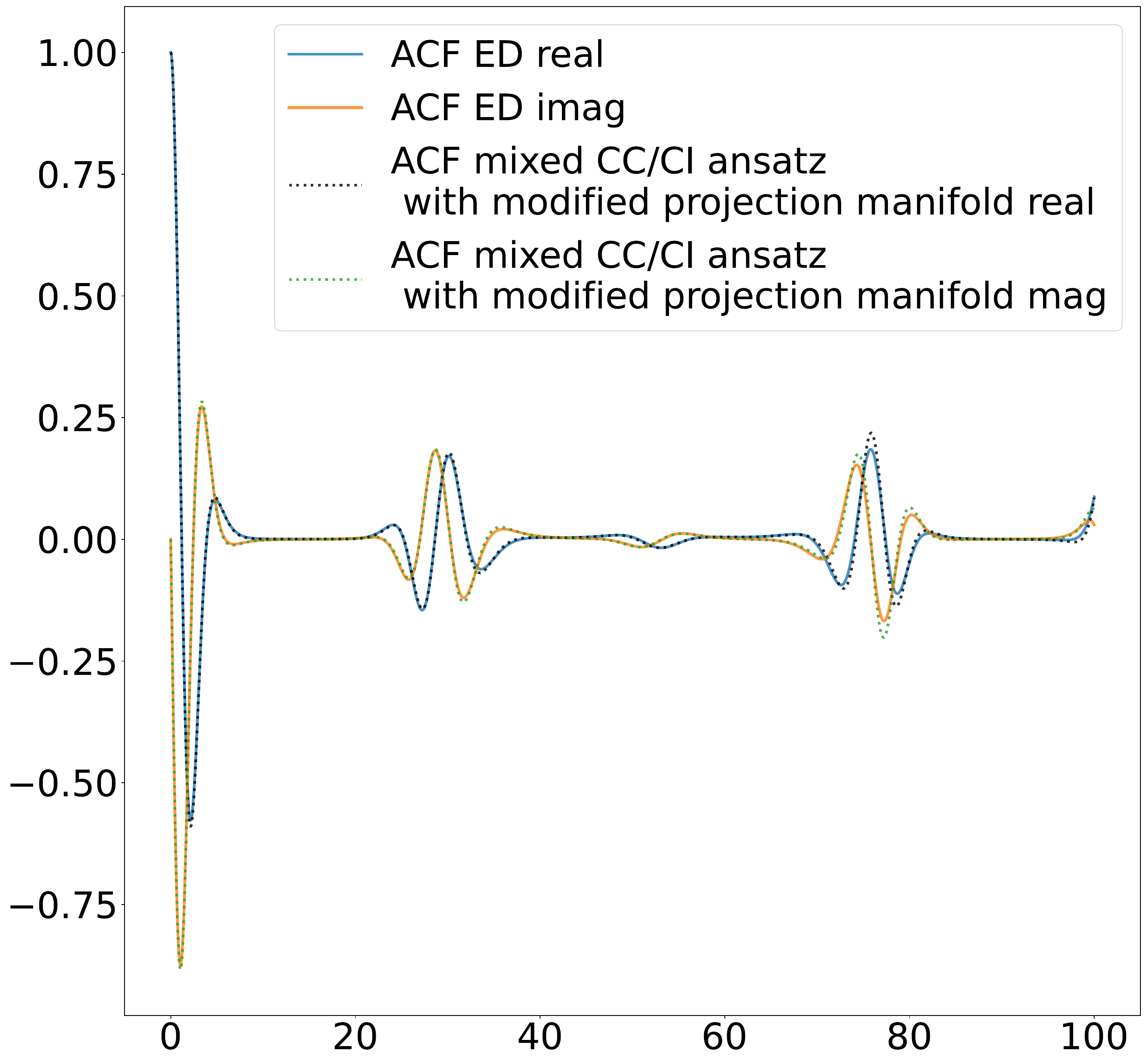}%
    }\hfill
    \subfloat[\label{subfig:four_b}%
        Absorption spectras of ED, mixed CC/CI ansatz, and mixed CC/CI ansatz with modified projection manifold.%
    ]{%
        \includegraphics[width=0.8\columnwidth]{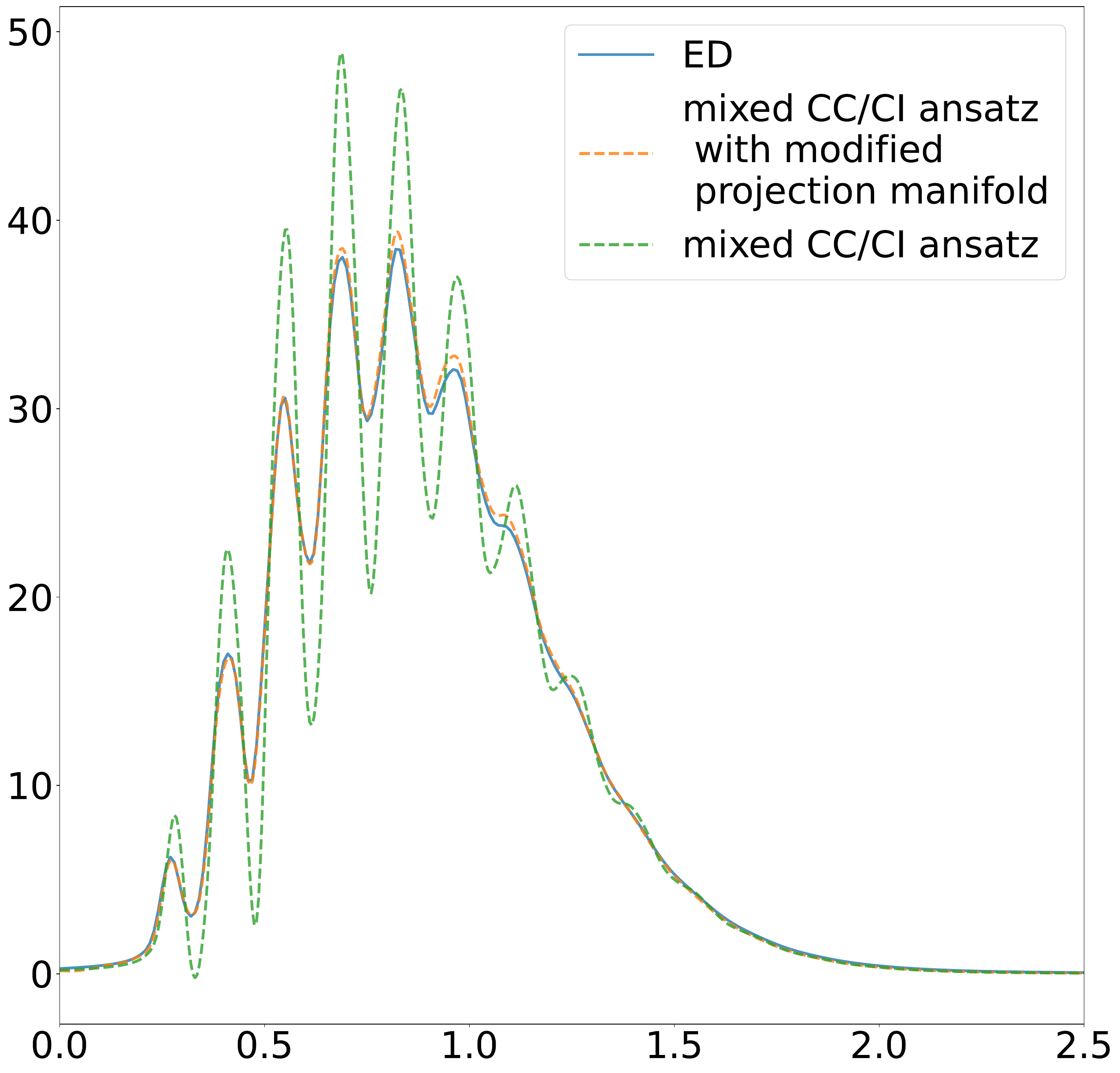}
    }\hfill
    \caption{%
        \label{fig:four}
        Introducing the modification of the projection manifold $\eTd$, further improves the numerical results of the time-autocorrelation function and the Fourier transformed spectra.
    }
\end{figure}

It is gratifying that modifying the projection manifold using $\eTd$ notably improves agreement with ED as seen in~\Fref{fig:four}.
Interestingly, the formal parameterization of the wave-function has not improved, but the values for the parameters have improved.
The modified projection manifold provides a moving reference that incorporates the linear displacement of the initial Gaussian wave-packet situated at the equilibrium position of ground state PES, and hence we project against states that reflect the current position of the wave packet.
If one does not adjust the projection manifold one may get ill defined equations for the time-derivatives, and they can easily start to diverge in multistate problems.
This is the situation one needs to avoid, and adjustment of the projection manifold to capture the position of the wave packet appears to be effective, and robust.

The scheme described in this section will be adopted and extended to the full vibronic models.
In a full vibronic model, when introducing multiple electronic surfaces, the partition of $\hat{T}$ and $\hat{Z}$ residual becomes more complicated.
In the following sections, we will describe how to tackle these complications and optimize the choice of the partitioning of $\hat{T}$ residual.

%
%
%

\subsection{\label{sec:Two_Two} Approach for full vibronic models}
In this section we develop the VECC parameterization for general multi-surface vibronic models, and we will discuss the detailed equations of motion for all the parameters.
Given a vibronic model Hamiltonian, as described in \Eref{eqn:vibronic_H}, our goal is to develop a computational method that can efficiently simulate the non-adiabatic dynamics of the model, with a focus on the autocorrelation function.
In the following sections, we will discuss how to extend the robust mixed CC/CI ansatz (defined in \Sref{sec:Two_One_Four}) from single-surface vibrational models to full vibronic models.
We outline a straightforward extension in \Sref{sec:Two_Two_One}.
Then, we improve on this extension using an Ehrenfest parameterization in \Sref{sec:Two_Two_Two}.
Finally, we present additional manipulations in \Sref{sec:Two_Two_Three}, which result in a significantly simplified computational procedure for the propagation of the amplitudes.

\subsubsection{\label{sec:Two_Two_One} Extending the single-surface approach to multiple surfaces}

The full wave-function of non-adiabatic vibronic models can be expressed as a linear combination of electronic states, which are each initiated in a single diabatic electronic state $b$
\begin{equation}\label{eqn:full_wfn}
    \ket{\Psi(\tau)} = \lsum{b}\chi_b\ket{\psi(\tau)}_b
\end{equation}
where the weight of each electronic state $\chi_b$ is the transition dipole moment, which we assume to be independent of geometry, as is reasonable for vibronic models.

At $\tau=0$ we assume, for each electronic surface, that all vibrational modes are in their ground vibrational state
\begin{equation}\label{eqn:initial_wfn}
    \ket{\psi(\tau=0)}_b = \ket{b, 0},
\end{equation}
where the labels $b$ and $0$, in the RHS ket $\ket{b, 0}$, denote electronic surface $b$ and ground vibrational levels (using GS normal modes) respectively.
The initial state of the
quantum dynamics propagation 
can then be defined as:~\footnote{%
This initial condition can be further generalized to adapt to different experimental setups, but these generalizations are beyond the scope of this paper.
}
\begin{equation}
    \ket{\Psi(\tau=0)} = \lsum{b} \chi_b \ket{b,0}.
\end{equation}

The wave-function of each electronic state, $b$ is parameterized using the mixed CC/CI ansatz
\begin{equation}\label{eqn:vib_ansatz}
    \ket{\psi(\tau)}_b = \lsum{x}e^{\hat{T}(\tau)}\hat{Z}_{x}(\tau)\ket{x, 0},
\end{equation}
where
\begin{equation}\label{eqn:vib_ansatz_p1}
    \hat{T}=\lsum{i}t^i\cop{i},
\end{equation}
and
\begin{equation}\label{eqn:vib_ansatz_p2}
    \hat{Z}_{x}=z_x^0+\lsum{i}z_x^i\cop{i}+\frac{1}{2}\lsum{ij}z_{x}^{ij}\cop{i}\cop{j}.
\end{equation}
As in the case of the single-surface vibrational problem in \Eref{eqn:_mix_CC_CI_ansatz}, this ansatz consists of two components: $\hat{T}$ and $\hat{Z}$.
The $\hat{T}$ operator is truncated at singles and has an exponential parameterization.
Importantly it has no electronic-surface dependence.
The operator $\hat{Z}$ is truncated to singles and doubles and has a linear CI-like parameterization.
It has components specific for each diabatic electronic state. Let us emphasize that the time propagation for each state $\ket{\psi(\tau)}_b$ is independent. The time-correlation function combining the information from the individual state propagations and transition dipoles is assembled later.

Substituting \Eref{eqn:vib_ansatz} into the TDSE and projecting against $\bra{y,0} \proj \emT$ gives the following EOM:
\begin{multline}\label{eqn:vib_EOM}
    \bra{y,0}
        \proj
        \left(  i\dv{\hat{T}}{\tau}\hat{Z}_y
                + i\dv{\hat{Z}_y}{\tau}
        \right)
    \ket{y,0}
    =
\\
    \lsum{x}\bra{y,0}\proj
    \bar{\textbf{H}} \hat{Z}_{x}\ket{x,0}
,
\end{multline}
where
\begin{equation}\label{eqn:sim_tran_H_multi_surface}
    \bar{\textbf{H}}
    = \emT \hat{\textbf{H}}\eT
    = \left(\hat{\textbf{H}}\eT \right)_{\conntext} .
\end{equation}

In \Eref{eqn:sim_tran_H_multi_surface} we transform each individual electronic component $\hat{H}^a_b$ with the same purely vibrational operator $e^{\hat{T}}$ and the result is connected.

As in the case of the single-surface approach, we modify the projection manifold by inserting $\eTd$ to produce new EOM:
\begin{multline}\label{eqn:vib_EOM_proj}
    \bra{y,0}
        \proj
        \eTd
        \left(  i\dv{\hat{T}}{\tau}\hat{Z}_y
                + i\dv{\hat{Z}_y}{\tau}
        \right)
    \ket{y,0}
    =
\\
    \lsum{x}\bra{y,0}\proj
    \eTd \bar{H}^y_x \hat{Z}_{x}\ket{x,0}
.
\end{multline}

\begin{figure}[!h]\centering%
    \includegraphics[width=\columnwidth]{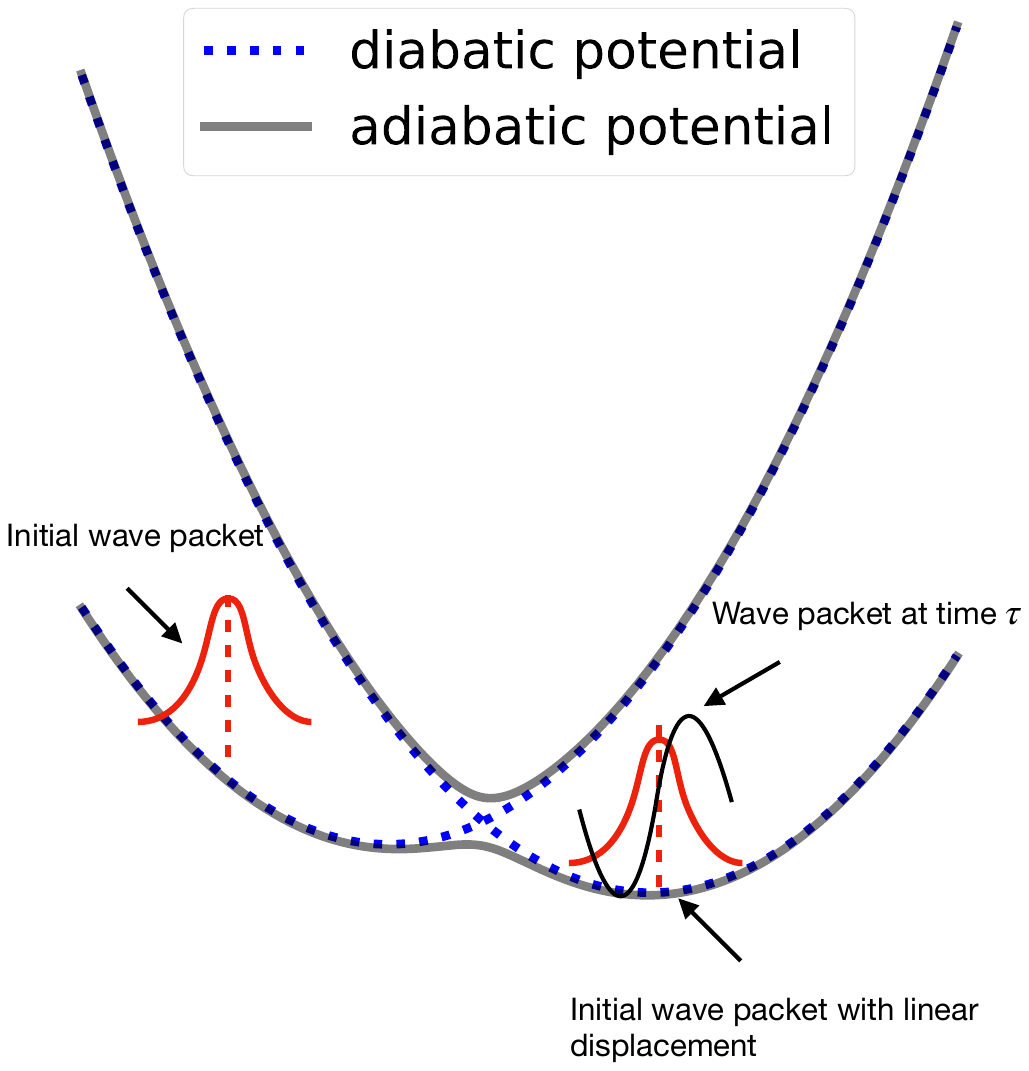}%
    \caption{%
        \label{fig:modify_proj}%
        By introducing $\eTd$, we shift the center of the initial wave-function, or reference state, to align it with the current wave-function at time $\tau$.
    }
\end{figure}

As illustrated in \Fref{fig:modify_proj}, the goal of introducing this modification of the projection manifold is the same as in the single-surface case.
It shifts the center of the reference state to align with the center of the moving wave-packet, thereby incorporating contributions of higher level excitations in the linear $\hat{Z}$ ansatz.
This modification improves both the accuracy and robustness.

Similar to the single-surface vibrational problem, the parameterization of the mixed CC/CI ansatz here is redundant.
One is free to choose how to partition the $\hat{T}$ and $\hat{Z}$ residuals from the ``net residual'' in the right hand side of \Eref{eqn:vib_EOM_proj}.
One straightforward choice is to parameterize $\hat{T}$ as the diagonal matrix elements of $\bar{H}$ (for the initial state $b$) as a direct analogy to the single-reference case (see \Eref{eqn:mix_CC_CI_EOM_2}):
\begin{equation}
i\dv{t^i}{\tau} = \bar{H}^i_{bb}.
\end{equation}
By definition the EOM for $\hat{Z}$ are the remaining components of~\Eref{eqn:vib_EOM_proj}.

We have therefore extended our mixed CC/CI ansatz from single-surfaces to general multi-surface vibronic models.
By evaluating the equations in this section we can efficiently simulate the non-adiabatic dynamics of the model.

While straightforward, this computational method has problems due to our partitioning of the $\hat{T}$ and $\hat{Z}$ residuals.
In \Eref{eqn:vib_ansatz}, the sum over $x$ is a sum over all electronic states.
$\hat{T}$ has no electronic label while $\hat{Z}$ has electronic surface dependence.
At this stage the $\hat{T}$ operator only captures the diagonal surface contributions from the initial state $b$, while the $\hat{Z}$ operator captures both the diagonal and off-diagonal contributions of the time dependent wave-function.
This is not ideal as the dominant contributions (the exponential operator which includes infinite fold excitations in the Taylor series expansion) do not incorporate the surface crossing contributions of the time dependent wave-function.

To tackle this, we devise an improved parameterization of the $\hat{T}$ (and therefore $\hat{Z}$) residuals in the following section.
This parameterization accommodates the switching of electronic surfaces of the moving wave-packet which, to some extent, allows $\hat{T}$ to incorporate surface-crossing contributions of the time-dependent wave-function.

\subsubsection{\label{sec:Two_Two_Two} Ehrenfest parameterization of \texorpdfstring{$\hat{T}$}{T} residuals}

The $\hat{T}$ component in the ansatz (\Eref{eqn:vib_ansatz}) has an exponential parameterization which incorporates high level excitations from its Taylor series expansion.
Therefore if the $\hat{T}$ component is the dominant contribution of the time-dependent wave-function we should expect more accurate results.
Our goal then, is to partition a large portion of the ``net residual'' on the $\hat{T}$ residual.
To achieve this goal, we borrow the idea of Ehrenfest dynamics.
Ehrenfest dynamics has been applied to ab-initio molecular dynamics, in which forces on nuclear motion are evaluated by taking weighted averages over all PES contributions when solving the Newtonian EOM~\cite{li2005ab}.

In our case, we parameterize the $\hat{T}$ residual as a ``weighted average'' over all electronic states, where the weight is the approximated CI coefficient of each electronic state in the time-dependent wave-function.
In this way, the $\eT$ component of the ansatz describes the ``average motion'' of the wave-packet over all electronic states.

Our Ehrenfest parameterization of the $\hat{T}$ residual is:
\begin{equation}\label{eqn:vib_EOM_T}
     i\dv{t^i}{\tau}= C \lsum{y} \left(c^{0}_{y}\right)^{*} R^{i}_{y},
\end{equation}
where $R^i_y$ is defined as the singles residual, containing only the constant order contributions of $\hat{Z}$:
\begin{equation}\label{eqn:singles_residual}
    R^i_y = \lsum{x} \bra{y,0} \aop{i}\eTd \bar{H} z_{x}^{0} \ket{x,0}.
\end{equation}
By setting the coefficients to be the zero order component of $\hat{Z}$ ($c^0_y = z^0_y$) they essentially parameterize the Hermitian conjugate of the zero order component of the wave-function:
\begin{equation}
    \bra{\psi_{0}} \approx \lsum{x}\bra{x,0} (z^0_x)^* = \lsum{x}\bra{x,0} (c^0_x)^*.
\end{equation}
In practice, our choice of the $c^0_y$ coefficients is different due to considerations discussed in the next section.
The coefficient $C$ is the real, positive normalization constant
\begin{equation}
    C = \left(\lsum{y} \left(c^{0}_{y}\right)^{*} c^{0}_{y}\right)^{-1}.
\end{equation}

By taking the electronic state average to parameterize the $\hat{T}$ residual, the surface crossing contribution of the time-dependent wave-function is incorporated.
For vibronic models with strong coupling, the wave-packet might be switched from its initial PES to another PES (or some linear combination) during the propagation.
The Ehrenfest dynamics parameterization of $\hat{T}$ residual can accommodate this change of surfaces through the weighted average over all electronic states.
When a switch occurs, the surface that the wave-packet switched to will have dominant weight in the time-dependent wave-function.

Using the Ehrenfest parameterization of $\hat{T}$, by substituting \Eref{eqn:vib_EOM_T} into \Eref{eqn:vib_EOM_proj}, we formulate new EOM for $\hat{Z}$
\begin{equation}\label{eqn:vib_EOM_Z}
    \begin{split}
        i\bra{y,0}
            \proj  \dv{\hat{Z}_y}{\tau}
        \ket{y,0}
        &=  \lsum{x} \bra{y,0}
            \proj  \eTd  \hat{\textbf{G}} \hat{Z}_{x}
        \ket{x,0}
    \\  &   -i\bra{y,0}
            \proj
            \left(
                \eTd -\identity
            \right)\dv{\hat{Z}_y}{\tau}
        \ket{y,0}
    \end{split}
\end{equation}
where
\begin{equation}
    \hat{G}\equiv\bar{\textbf{H}}-i\dv{\hat{T}}{\tau} \textbf{1}
\end{equation}

The formulation described above gives an efficient computational method to simulate wave-packet dynamics of non-adiabatic vibronic models.
By introducing the idea of Ehrenfest dynamics to partition the $\hat{T}$ residual, we have improved the choice of the $\hat{T}$ residual parameterization.
Because $\hat{T}$'s EOM (\Eref{eqn:vib_EOM_T}) are coupled to $\hat{Z}$'s EOM (\Eref{eqn:vib_EOM_Z}), both the $\hat{T}$ and $\hat{Z}$ components of the ansatz incorporate the vibronic coupling effects.

In the next section, we introduce a minor modification of the current scheme.
The motivation to introduce this modification is purely technical; the modification does not affect the end result, but rather simplifies the software implementation and increases the computational efficiency.

\subsubsection{\label{sec:Two_Two_Three} Simplification of the EOM through Similarity transformation}
As been discussed previously, the modified projection manifold can improve both robustness and accuracy of the VECC method.
However, there are some practical challenges with implementing the modification.
The $\eTd$ operator will dramatically increase the number of terms in the CC residual equations.
This has two side effects.
It increases the difficulty of manually deriving the CC residual equations and the corresponding EOM.
It also results in more terms being present in the CC residual equations, which makes the computational method more expensive.
In order to reduce the chance of human errors in deriving the working equations as well as reduce the computational cost, we introduce a modification of the scheme described in the last section.

We insert resolutions of identity $\left(e^{-\hat{T}^\dagger} \eTd \right)$ into \Eref{eqn:vib_EOM_proj}:
\begin{equation}\label{eqn:vib_STEOM_1}
    \begin{split}
        \bra{y,0}
        &   \proj  \eTd
            \left(
                i\dv{\hat{T}}{\tau} \left(e^{-\hat{T}^\dagger} \eTd \right) \hat{Z}_y
                + i\dv{\hat{Z}_y}{\tau}
            \right)
        \ket{y,0}
    =
    \\  &\lsum{x} \bra{y,0}
            \proj  \eTd  \bar{\textbf{H}}
            \left(e^{-\hat{T}^\dagger} \eTd \right)
            \hat{Z}_{x}
        \ket{x,0}
        .
    \end{split}
\end{equation}
We can simplify \Eref{eqn:vib_STEOM_1} by defining four new quantities
\begin{subequations}\label{eqn:ST_all}
    \begin{equation}
    \label{eqn:ST_H}
        \tilde{\bar{\textbf{H}}}
        \equiv  \eTd  \left(\bar{\textbf{H}}\right) e^{-\hat{T}^\dagger}
        =       \left(\eTd  \bar{\textbf{H}}\right)_{\conntext}
    \end{equation}\begin{equation}
    \label{eqn:ST_T}
        \hat{\rho}
        \equiv  \eTd  \left(i\dv{\hat{T}}{\tau}\right)e^{-\hat{T}^\dagger}
        =       \left(\eTd  i\dv{\hat{T}}{\tau}\right)_{\conntext}
    \end{equation}\begin{equation}
    \label{eqn:ST_Z}
        \hat{C}_{x} \ket{x,0}
        \equiv  \left(\eTd  \hat{Z}_{x} \right)\ket{x,0}
        =       \left(\eTd  \hat{Z}_{x}\right)_{\conntext} \ket{x,0}
    \end{equation}\begin{equation}
    \label{eqn:ST_D}
        \hat{D}_{y} \ket{y,0} \equiv \left(\eTd  - \identity\right) i\dv{\hat{Z}_y}{\tau} \ket{y,0}
    \end{equation}
\end{subequations}
that can be used to formulate more compact EOM for $\hat{Z}$:
\begin{equation}\label{eqn:vib_STEOM_Z}
    \begin{split}
        i\bra{y,0} \proj  \dv{\hat{Z}_{y}}{\tau} \ket{y,0}
    &=  \lsum{x} \bra{y,0} \proj  \left( \tilde{\bar{\textbf{H}}} - \hat{\rho} \right) \hat{C}_{x} \ket{x,0}
\\  &-  \bra{y,0} \proj  \hat{D}_{y} \ket{y,0}.
    \end{split}
\end{equation}

All four quantities can be regarded as similarity transformations, although the last two may not be immediately obvious. It uses $e^{\hat{T}^{\dagger}} \ket{0} = \ket{0}$.
In addition, $\tilde{\bar{\textbf{H}}}$ can be regarded as a doubly-transformed Hamiltonian, first by $\eT$ from the  right, then by $\eTd$ from the left.
Another detail is that $\hat{C}_x$ from \Eref{eqn:ST_Z} has the form $c^0_x + \hat{c}^1_x + \hat{c}^2_x$ where the constant term $c^{0}_{x}$ defines the weights of electronic states in the Ehrenfest parameterization of $\hat{T}$ amplitude in \Eref{eqn:vib_EOM_T}.
To numerically integrate the cluster amplitudes we pre-calculate all four intermediate quantities and then use them to calculate the $\hat{Z}$ residuals, as seen in \Eref{eqn:vib_STEOM_Z}.

This new scheme has two main advantages.
The first is that we simplify the EOM of $\hat{Z}$.
\Eref{eqn:vib_STEOM_Z} resembles a linear CI equation which has fewer terms in its residual than the exponential form in \Eref{eqn:vib_EOM_Z}.
Consequently, the chance for mistakes when manually deriving the working equations of $\hat{Z}$ is reduced.
Secondly, by pre-calculating the intermediate quantities in \Eref{eqn:ST_all} we avoid repeatedly evaluating them in the CC residuals, which reduces the computational cost compared to the original scheme.

It is important to highlight that the purpose of this similarity transformation scheme is purely technical.
The new scheme is mathematically equivalent to the previous scheme and both will produce identical results.
In our opinion, this new concise formulation of VECC theory also bears some aesthetic value.
In this way, we have finalized the formal VECC theory for non-adiabatic nuclear dynamics.
In the next section we will discuss how to compute time dependent properties using  VECC theory.

\subsection{\label{sec:Two_Three} Calculating time-dependent properties with VECC}

As previously described, the general procedure for calculating time-dependent properties using VECC is fourfold: %
1.~the time-dependent wave-function is parameterized using an ansatz; %
2.~EOM are formed by substituting that ansatz into the TDSE; %
3.~cluster amplitudes are obtained by solving these EOM numerically.
4.~properties of interest can be mapped from the cluster amplitudes.

In~\Sref{sec:Two_Two_One} we parameterized the time-dependent wave-function using the mixed-CC/CI ansatz.
In the previous two sections we outlined the EOM of interest and the expressions for the $\hat{T}$ and $\hat{Z}$ (cluster) amplitudes.
In this section we describe the procedure for numerically evaluating those amplitudes and how they are mapped to properties of interest.

We specify the initial conditions of the amplitudes:
\begin{subequations}
    \begin{equation}\label{eq:initial_condition_T}
        \hat{T}(\tau=0) = \textbf{0},
    \end{equation}\begin{align}\label{eq:initial_condition_Z}
        \hat{Z}_{x}(\tau=0) &= \delta_{x, b},
    \end{align}
\end{subequations}
where the surface label $b$ denotes the electronic surface of interest.
Time-integration of the $\hat{T}$ and $\hat{Z}$ amplitudes (as given in
\Eref{eqn:vib_EOM_T} and \Eref{eqn:vib_STEOM_Z}) is performed one surface ($b$) at a time.
Because these amplitudes parameterize the time-dependent wave-function $\ket{\Psi(\tau)}$, time-dependent properties can be mapped through the amplitudes.
We will first look at the time-autocorrelation function (ACF).

\subsubsection{\label{sec:Two_Three_One} Calculation of time-autocorrelation function}

To compute the time-autocorrelation function (ACF), we first construct the matrix elements of the quantum propagator
\begin{equation}
    U_{ab} \equiv \bra{a,0}e^{-i\hat{H}\tau}\ket{b,0},
\end{equation}
in a ``column by column'' fashion.
Each column $b$ is obtained by numerical time integration starting from it's initial conditions (as specified in \Erefpl{eq:initial_condition_T}{eq:initial_condition_Z}).
\begin{equation}\label{eqn:map_U}
    U_{ab} = \bra{a,0} e^{\hat{T}} \sum_y \hat{Z}_y \ket{y,0} = z^0_a \quad \text{for } b \in \left\{b_{1}, b_{2}, \cdots \right\}
\end{equation}
By substituting the expression of the $\textbf{U}$ matrix from \Eref{eqn:map_U} into the definition of the ACF in \Eref{eqn:acf_definition}, we can map it to the CC amplitudes:
\begin{equation}\label{eqn:vib_ACF_1}
    \begin{split}
        C(\tau)
        &=      \bra{\Psi(t=0)} \ket{\Psi(t=\tau)}
    \\  &=      \lsum{ab} \chi_{a} \bra{a,0} e^{-i\hat{H}\tau} \ket{b,0} \chi_{b}
    \\  &\equiv \lsum{ab} \chi_{a} U_{ab}(\tau) \chi_{b},
    \end{split}
\end{equation}
where $\chi_b$ denotes the transition dipole moment of the diabatic electronic state $b$.
Thus, by calculating $U_{ab}(\tau)$ we can straightforwardly obtain the ACF for a system of interest.

\subsubsection{\label{sec:Two_Three_Two} General expression for expectation value \texorpdfstring{$\expval{O(\tau)}$}{} }
In general, the time-dependent expectation value for a quantum operator $\hat{O}$ is defined as
\begin{equation}\label{eqn:property_general_1}
    \expval{O(\tau)}
    = \frac{
        \expval{\hat{O}}{\psi(\tau)}
    }{
        \braket{\psi(\tau)}
    }.
\end{equation}
The evaluation of this expression for a general operator is nontrivial.
The first challenge is that in conventional coupled-cluster theory for electronic structure, the cluster expansion series for both the numerator and denominator are formally infinite (when directly applying the CC ansatz into them).
Another concern is that the CC amplitudes are not small in general.
However, in VECC theory we use a series of mathematical manipulations to evaluate terms without truncation, and in a simple manner.
The essential idea is to introduce a similarity transformation to cancel out the disconnected contributions in the expression, resulting in a finite expansion of the CC amplitudes.
In this section, we derive a mathematical expression for a general time-dependent property as given by \Eref{eqn:property_general_1} in a concise form (finite expansion).
\par
We begin by evaluating the denominator in \Eref{eqn:property_general_1}, using the VECC ansatz from \Eref{eqn:vib_ansatz}:
\begin{equation}\label{eqn:norm_exp_1}
    \braket{\psi(\tau)} =  \lsum{x,y}
    \bra{y,0}
        \hat{Z}_{y}^{\dagger} \eTd \eT \hat{Z}_{x}
    \ket{x,0}.
\end{equation}
Let us note here that we assume a single propagated wave function in this section, and do not propagate a superposition of states starting from an initial state for each surface. This simplifies the procedure and serves our current purposes.
We insert a resolution of the identity
\begin{equation}\label{eqn:norm_exp_2}
    \braket{\psi(\tau)} = \lsum{x}
    \bra{x,0}
        \hat{Z}_{x}^{\dagger}
        \left(\eT \emT \right) \eTd \eT
        \hat{Z}_{x}
    \ket{x, 0},
\end{equation}
resulting in the term $\emT \eTd \eT$, which we can evaluate by expanding using a Baker-Campbell-Hausdorf (BCH) series
\begin{equation}
    e^{\hat{s}_{1}} \equiv \emT \eTd \eT
    \quad \text{where} \quad
    \left(\hat{T}\equiv\lsum{i}t^{i}\cop{i}\right),
\end{equation}
such that
\begin{equation}\label{eqn:BCH_exp_eT_dagger}
    \begin{split}
    \hat{s}_1
        &=  \hat{T}^\dagger
            + \left[\hat{T}^\dagger, \hat{T}\right]
            + \frac{1}{2!}\left[\left[\hat{T}^\dagger,\hat{T}\right],\hat{T}\right]
            + \cdots
    \\  &=  \lsum{i}(t^i)^*\aop{i} + s_0
    \\  &=  \hat{T}^{\dagger} + s_0
            \quad \text{where} \quad
            s_0 \equiv \lsum{i} (t^i)^{*} t^i
    \end{split}
\end{equation}
Notice that this (infinite order) formula is particularly simple when using $\hat{T}$ truncated at singles and this is the prime reason that we use only singles in the $\hat{T}$ operator.

Substituting the BCH series from \Eref{eqn:BCH_exp_eT_dagger} into \Eref{eqn:norm_exp_2}, results in
\begin{equation}\label{eqn:norm_exp_3}
    \begin{split}
        \braket{\psi(\tau)}
        &=      \lsum{x}\bra{x,0}\hat{Z}_{x}^\dagger \eT \left(\eTd e^{s_0}\right) \hat{Z}_{x}\ket{x,0}
    \\  &\equiv e^{s_0}\lsum{x}\bra{x,0}\hat{Z}_{x}^\dagger \eT \eTd \hat{Z}_{x}\ket{x,0}.
    \end{split}
\end{equation}
We continue by inserting more resolutions of identity
\begin{equation}\label{eqn:denominator_1}
    e^{s_0}\lsum{x}\bra{x,0}\hat{Z}_{x}^\dagger
    \eT  \left(\lsum{n} \ket{x,n}\bra{x,n}\right)
    \eTd \hat{Z}_{x}\ket{x,0}
\end{equation}
such that
\begin{equation}\label{eqn:denominator_2}
    \braket{\psi(\tau)} \equiv e^{s_0} \lsum{x}\lsum{n} C_{n,x}^\dagger C_{n,x}
\end{equation}
where
\begin{equation}
    C_{n,x} \equiv \bra{x,n} \eTd  \hat{Z}_{x} \ket{x,0}.
\end{equation}
We can easily expand \Eref{eqn:denominator_1} by applying Wick's theorem.
In this fashion we have obtained an expression for the denominator as a finite expansion of the CC amplitudes.
Next we apply similar mathematical tricks to evaluate the numerator.

We begin by substituting the VECC ansatz (from~\Eref{eqn:vib_ansatz}) into the expression of the numerator in \Eref{eqn:property_general_1} as follows:
\begin{equation}\label{eqn:numerator_1}
    \bra{\psi(\tau)}\hat{O}\ket{\psi(\tau)}
    = \lsum{x,y}\bra{y,0}\hat{Z}_{y}^{\dagger} \eTd \hat{O} \eT \hat{Z}_{x}\ket{x,0}.
\end{equation}
We introduce similarity transformations (in the same fashion as we did previously) by inserting a resolution of the identity, transforming $\eTd \hat{O} \eT$ into $ \eTd \eT \bar{O}$
where
\begin{equation}
    \bar{O} \equiv
        \emT  \hat{O} \eT
    =   \left(\hat{O} \eT \right)_{\conntext}.
\end{equation}
Following that, $\eTd \eT \bar{O}$ is replaced with $\eT \eTd e^{s_0} \bar{O}$ by inserting another resolution of the identity and using the same BCH expansion from \Eref{eqn:BCH_exp_eT_dagger} ($\emT \eTd \eT \rightarrow \eTd e^{s_0}$).
In this manner we can re-express \Eref{eqn:numerator_1} as:
\begin{equation}\label{eqn:O_exp_1}
    \begin{split}
        \bra{\psi(\tau)}\hat{O}\ket{\psi(\tau)}
        &=      \lsum{x,y} \bra{y,0}\hat{Z}_{y}^{\dagger} \eT \eTd e^{s_0} \bar{O}\hat{Z}_{x}\ket{x,0}
    \\  &\equiv e^{s_0} \lsum{x,y} \bra{y,0}\hat{Z}_{y}^{\dagger} \eT \eTd \bar{O}\hat{Z}_{x}\ket{x,0}.
    \end{split}
\end{equation}
Then, as in \Eref{eqn:denominator_1}, we insert an additional resolution of identity:
\begin{equation}\label{eqn:O_exp_2}
    e^{s_0}\lsum{x,y}
    \bra{y,0}
        \hat{Z}_{y}^\dagger \eT
        \left(\lsum{n} \ket{y,n}\bra{y,n}\right)
        \eTd \bar{O}\hat{Z}_{x}
    \ket{x,0},
\end{equation}
such that
\begin{equation}\label{eqn:O_exp_3}
    \bra{\psi(\tau)}\hat{O}\ket{\psi(\tau)}
    =   e^{s_0} \lsum{x,y,n} C^{\dagger}_{n,y}
        \bra{y,n}
            \eTd \bar{O} \hat{Z}_{x}
        \ket{x,0}.
\end{equation}
where the term $\bra{y,n} \eTd  \bar{O}\hat{Z}_{x} \ket{x,0}$ in \Eref{eqn:O_exp_3} has the explicit form of the right hand side in the CC EOM from \Eref{eqn:vib_EOM_proj}.
We can evaluate this term using VECC residual equations of the form in \Eref{eqn:singles_residual} simply by substituting the matrix elements of $\bar{O}$ in place of $\bar{H}$.

Thus, we have derived an expression for both the denominator and numerator of a general time-dependent property, as expressed in \Eref{eqn:property_general_1}, with finite truncation of the VECC amplitudes.

We will further simplify \Eref{eqn:O_exp_3}, by applying the similarity transformation scheme discussed in \Sref{sec:Two_Two_Three}.
We start by inserting resolutions of the identity:
\begin{equation}
    \begin{split}
        &   \bra{\psi(\tau)} \hat{O} \ket{\psi(\tau)}
    \\  &=   e^{s_0} \lsum{x,y,n} C^{\dagger}_{n,y}
        \bra{y,n}
            \eTd \bar{O}
            \left(e^{-\hat{T}^\dagger}e^{\hat{T}^{\dagger}}\right)
            \hat{Z}_{x}
        \ket{x,0}.
    \end{split}
\end{equation}
As before, we define similarity transformed intermediate quantities (in this case we only define the two quantities for the RHS):
\begin{subequations}
    \begin{equation}
        \tilde{\bar{O}}
        \equiv  \eTd \bar{O}e^{-\hat{T}^\dagger},
    \end{equation}\begin{equation}
        \hat{C}_{x}\ket{x,0}
        \equiv  e^{\hat{T}^{\dagger}}\hat{Z}_{x}\ket{x,0}
        =       \left(\eTd \hat{Z}_{x}\right)_{\conntext}\ket{x,0}.
    \end{equation}
\end{subequations}
Substituting these intermediate quantities into \Eref{eqn:O_exp_3}
\begin{equation}\label{eqn:O_exp_4}
    \bra{\psi(\tau)}\hat{O}\ket{\psi(\tau)}
    =   e^{s_0} \lsum{x,y,n} C^{\dagger}_{n,y}
        \bra{y,n} \tilde{\bar{O}} \hat{C}_{x} \ket{x,0} .
\end{equation}
Following the arguments in~\Sref{sec:Two_Two_Three} by pre-calculating the intermediate quantities, one can simplify~\Eref{eqn:O_exp_3} and reduce the computational cost.
Additionally, \Eref{eqn:O_exp_4} is mathematically equivalent to \Eref{eqn:O_exp_3} and the expectation value is invariant with respect to the insertion of the resolution of the identity.

Substituting the numerator expression in \Eref{eqn:O_exp_4} and the denominator expression in \Eref{eqn:denominator_2} into \Eref{eqn:property_general_1} we can express the expectation value of a generic time-dependent property (in a finite expansion of the VECC amplitudes) as follows:

\begin{align}\label{eqn:property_general_2}
    \expval{\hat{O}(\tau)}
    &= \frac{
        \lsum{x,y,n} C^{\dagger}_{n,y}
        \bra{y,n} \tilde{\bar{O}} \hat{C}_{x} \ket{x,0}
    }{
        \lsum{x,n} C_{n,x}^\dagger C_{n,x}
    }.
\end{align}
Thus, we derived an expression in a concise form.

In the next section we will focus on calculating diabatic state populations, although in general the scheme discussed in this section can be applied to calculate an arbitrary expectation value of time-dependent properties.\@
The implementation of approaches to calculate time-dependent properties other than the diabatic state population is beyond the scope of this paper, but it opens the possibilities for future work.

\subsubsection{\label{sec:Two_Three_Three} Calculation of diabatic state population}
If we are interested in a specific property (e.g. the diabatic state population) we simply choose a form for the operator $\hat{O}$
\begin{equation}
    \hat{O} \equiv \lsum{n} \ket{a,n}\bra{a,n},
\end{equation}
and substitute it into our concise form in \Eref{eqn:property_general_2} resulting in an expression for the diabatic state population:
\begin{equation}\label{eqn:state_pop}
    \begin{split}
        P_a(\tau)
        = \frac{
            \lsum{n} C^\dagger_{n,a} C_{n,a}
        }{
            \lsum{x,n} C_{n,x}^\dagger C_{n,x}
        }.
    \end{split}
\end{equation}
In this way, we have derived a concise expression for the diabatic state population.
In our implementation of VECC we use \Eref{eqn:state_pop} to calculate the diabatic state population.
In practice, we begin by determining the $\hat{T}$ and $\hat{Z}$ amplitudes through numerical real-time integration.
At each time step, the diabatic state population is mapped from the $\hat{T}$ and $\hat{Z}$ amplitudes in an ``on-the-fly'' fashion.

\section{\label{sec:Three} Result and discussion}
The VECC approach (described in \Sref{sec:Two_Two}) is implemented in the open source code \texttt{t-amplitudes}.~\cite{T-amplitudes}
We evaluate the VECC method at both doubles (SD) and triples (SDT) truncation of the $\hat{Z}$ amplitude and with singles (S) truncation of the $\hat{T}$ amplitude.
The explicit working equations used in \texttt{t-amplitudes} are generated by the equation generator developed by Raymond et.\ al.\ , whose open source software package \texttt{termfactory} is available on GitHub.~\cite{raymondtermfactory2022}
In the implementation, CC amplitudes are obtained through numerical real-time integration, and time-dependent properties are mapped from those CC amplitude.
Third-party Python libraries NumPy and SciPy are utilized in this implementation to solve numerical ODEs and tensor products.~\cite{GitNumPy,GitSciPy}
The code is optimized to reach high performance.

We simulated the vibrationally-resolved electronic spectra in \Sref{sec:Three_One}, and the diabatic state population in \Sref{sec:Three_Two}.
All results were computed using the implementation of the VECC method as described in \Sref{sec:Two_Three}.
These results were benchmarked against MCTDH and exact diagonalization (ED).
\subsection{\label{sec:Three_One} Vibrationally resolved electronic absorption spectra}

\NewDocumentCommand{\modelA}{}{model-A}
\NewDocumentCommand{\modelB}{}{model-B}
\NewDocumentCommand{\tbyt}{}{$2$-by-$2$}

In \Sref{sec:Three_One_One} we explore representative vibronic spectra of small \tbyt{} models (two electronic surfaces and two normal modes).
Since \tbyt{} models have a small number of degrees of freedom, we can afford to compare VECC results against ED in a converged basis set.
This allows for convenient testing of the VECC method for spectra simulation for a variety of vibronic models.
The model systems we discuss here are designed to mimic strong vibronic coupling scenarios.

In \Sref{sec:Three_One_Two} we investigate the photo-electron spectra of ``small'' molecules, specifically water (\ce{H2O}), and ammonia (\ce{NH3}).
For these models we anticipate good agreement with the (converged) MCTDH results.
The goal of this section is to demonstrate the validity of the VECC method for real molecular systems.

Then in \Sref{sec:Three_One_Three} we look at ``larger'' molecules, specifically the widely-tested pyrazine (\ce{C4H4N2}), and hexahelicene (\ce{C26H16}).
In these complicated models we are more focused on qualitative comparison with the MCTDH results.
The intention is to highlight the efficiency of the VECC method, while capturing the qualitative spectra.

\par
The \tbyt{} models are described by the Hamiltonian:
\begin{equation}\label{eqn:two_by_two_model}
    \begin{split}
        \hat{H} &= \hat{K} + \hat{V}
    \\  &=
        \begin{bmatrix}
            E_a + h_0 + \lambda_1 q_1 & 0
        \\  0 & E_b + h_0 + \lambda_2 q_1
        \end{bmatrix}
        +
        \begin{bmatrix}
            0 & \gamma q_2
        \\ \gamma q_2  & 0
        \end{bmatrix}
    \\  & h_0 = \frac{1}{2}\Big(\omega_{1} q_{1}^{2} + \omega_{2} q_{2}^{2}\Big),
    \end{split}
,
\end{equation}
where $E_a$ and $E_b$ are vertical energies corresponding to the two electronic states, $\omega_1$ and $\omega_2$ are the frequencies for the two vibrational modes, $\lambda_1$ and $\lambda_2$ are linear displacements corresponding to the two PESs and $\gamma$ dictates the strength of the Peierls (off-diagonal) vibronic couplings.
Of the various \tbyt{} models we tested we will explicitly show two, which we will call \modelA{} and \modelB{}.
\begin{table}[h!]\centering
    \caption{\label{table:one}\tbyt{} model parameters.}
    \begin{tabular}{
        c
        S[
            table-figures-integer = 1,
            table-figures-decimal = 3,
            table-number-alignment = center
        ]
    }
        \toprule
        Parameter & {(\si{\electronvolt})} \\
        \midrule
        $E^{a}$      & 0.0 \\
        $E^{b}$      & 0.2 \\
        $\omega_{1}$ & 0.2 \\
        $\omega_{2}$ & 0.3 \\
        $\gamma$     & 0.2 \\
        \botrule
    \end{tabular}
\end{table}

\begin{table}[h!]\centering
    \caption{\label{table:two} in (\si{\electronvolt}).}
    \begin{tabular}{
        c
        S[
            table-figures-integer = 1,
            table-figures-decimal = 3,
            table-number-alignment = center
        ]
        S[
            table-figures-integer = 1,
            table-figures-decimal = 3,
            table-number-alignment = center
        ]
    }
        \toprule
        Parameter & {\modelA{}} & {\modelB{}}\\
        \midrule
        $\lambda_{1}$  &   0.5  &  1.0 \\
        $\lambda_{2}$  &  -0.25 & -0.5 \\
        \botrule
    \end{tabular}
\end{table}
Both models use the parameters given in~\Tref{table:one}.
\modelA{} and \modelB{} only differ in their linear displacement parameters $\lambda_1$ and $\lambda_2$.
These models are typical strong vibronic models since they have a small energy gap between the two electronic states and sizeable Peierls vibronic-coupling constants.
In \modelA{}, we set $\lambda_1 = \SI{0.5}{eV}$ and $\lambda_2 = \SI{-0.25}{eV}$.
This model has relatively smaller linear displacements which resembles the short progression case for the molecular spectra.
In \modelB{}, we set $\lambda_1 = \SI{1.0}{eV}$ and $\lambda_2 = -\SI{0.5}{eV}$, which mimics the long progression scenario.

The water and ammonia models were generated using code developed by the Nooijen group.
Constructing these models begins by obtaining excited states energies.
This was accomplished using electronic structure calculations, specifically an IP-EOMCCSD method.~\cite{%
bomble2005equation,musia2004eom,musial2003equation,stanton1994analytic,nooijen1993coupled,nooijen1992coupled,bartlett1994reviews}
Next, a diabatization procedure was applied using the VIBRON software package.~\cite{nooijen2003vibron}
These calculations were performed by Julia Endicott.~\cite{%
julia2014chem494report,marcel2011modelparameters}

The water model~(3 states, 3 modes) is a typical weak-coupling system.
The energy gap between each electronic state is very large $\sim$ \SI{3}{eV}.
The vibronic couplings are dominated by Holstein couplings (diagonal matrix elements) and there is only one Peierls coupling (off-diagonal elements).

The ammonia model~(3 states, 6 modes) is a typical strong-coupling system.
Two of the three states are degenerate Jahn-Teller states whose energy levels are higher than the non-degenerate state.
All of the three states are coupled with each other through Peierls couplings.
While the two degenerate Jahn-Teller states are strongly coupled with each other, the non-degenerate state is weakly coupled with the two degenerate Jahn-Teller states due to the large energy separation ($\sim$ \SI{6}{eV}) between them.

The pyrazine and hexahelicene models were obtained from the literature.
The widely-tested pyrazine model was obtained from the reference by Raab et al.~\cite{raab1999molecular}
This model has 2 electronic states and 24 vibrational modes.
The two electronic states have $\sim$ \SI{1}{eV} energy separation.
The vibronic-coupling constants are truncated to bilinear couplings.
The model parameters, time of propagation, and settings of the Fourier transform are obtained from the reference.

The hexahelicene model was obtained from the reference by Aranda et al.~\cite{aranda2021vibronic}
This model has 14 electronic states and 63 vibrational modes.
All electronic states in this model are close lying, with $\sim$ \SI{0.2}{eV} energy separation.
Vertical energies were computed using TD-DFT with only linear vibronic-coupling constants (LVC).
We benchmarked our VECC calculation against the multi-layer (ML)-MCTDH calculation provided in the reference.

\subsubsection{\label{sec:Three_One_One}Spectra for two surfaces two modes model systems}
\begin{figure}[!h]
    \subfloat[\label{subfig:small_lambda}%
        Spectra of \modelA{} propagated for 100 fs.
    ]{
        \includegraphics[width=0.8\columnwidth]{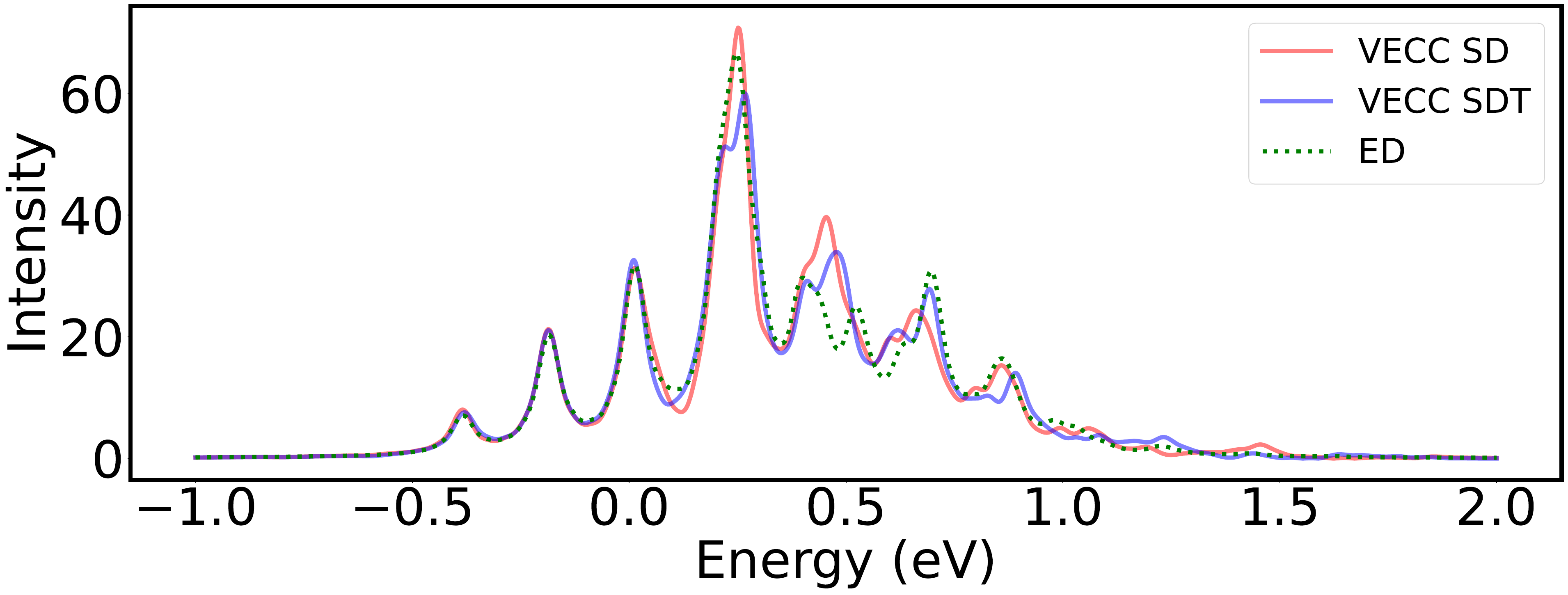}%
    }\hfill
    \subfloat[\label{subfig:large_lambda}%
        Spectra of \modelB{} propagated for 100 fs.
    ]{%
        \includegraphics[width=0.8\columnwidth]{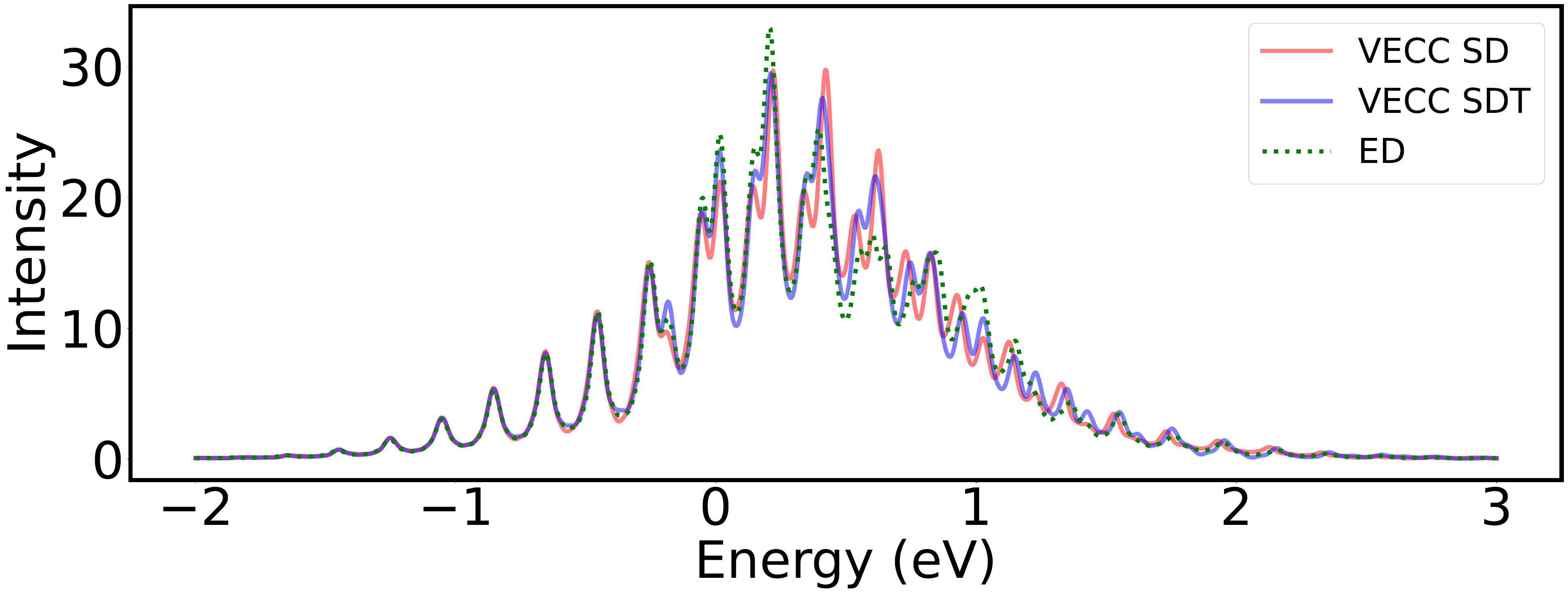}
    }\hfill
    \caption{%
        \label{fig:two_by_two_models}
        Simulated absorption spectra of \tbyt{} models using VECC and ED.
        Both are generated through a Fourier transform of their time-autocorrelation functions using~\texttt{autospec84}.
        Parameters for both models are given in \Tref{table:one} and \Tref{table:two}.
    }
\end{figure}

To illustrate the accuracy of VECC we compare to ED for the \tbyt{} models, which are typical of strong-coupling systems.
Results are shown in~\Fref{fig:two_by_two_models}.

The simulated spectra for~\modelA{} is given in~\Fref{subfig:small_lambda}.
This is a typical system with a short progression.
In general, VECC has good agreement with ED.
There are small discrepancies in terms of fine-structure but the band width and shape roughly matches.
The spectra for triples (SDT) truncation shows only slight differences with the spectra for doubles (SD) truncation, indicating that VECC is sufficiently converged at SD.

The simulated spectra for~\modelB{} is given in~\Fref{subfig:large_lambda}.
This is a typical system with long progression.
As expected, the simulated spectra shows richer peaks compared to the short progression case.
VECC predicts peak positions and intensities consistent with ED.
The spectra for SDT and SD show slight differences indicating that VECC is sufficiently converged at SD.

The VECC method shows impressive accuracy.
The envelope of the spectra (which is determined by very-short time evolution) is always well reproduced.
Somewhat surprisingly, in the ED method, a large basis set (80 basis functions per mode) is required to obtain converged simulated spectra which is computationally expensive and not always feasible for large systems.

\subsubsection{\label{sec:Three_One_Two} Spectra for small molecule compounds}

\begin{figure}[!h]
    \subfloat[\label{subfig:water_spectra}%
        Photo-electron spectra of water after 100 fs propagation.%
    ]{
        \includegraphics[width=0.8\columnwidth]{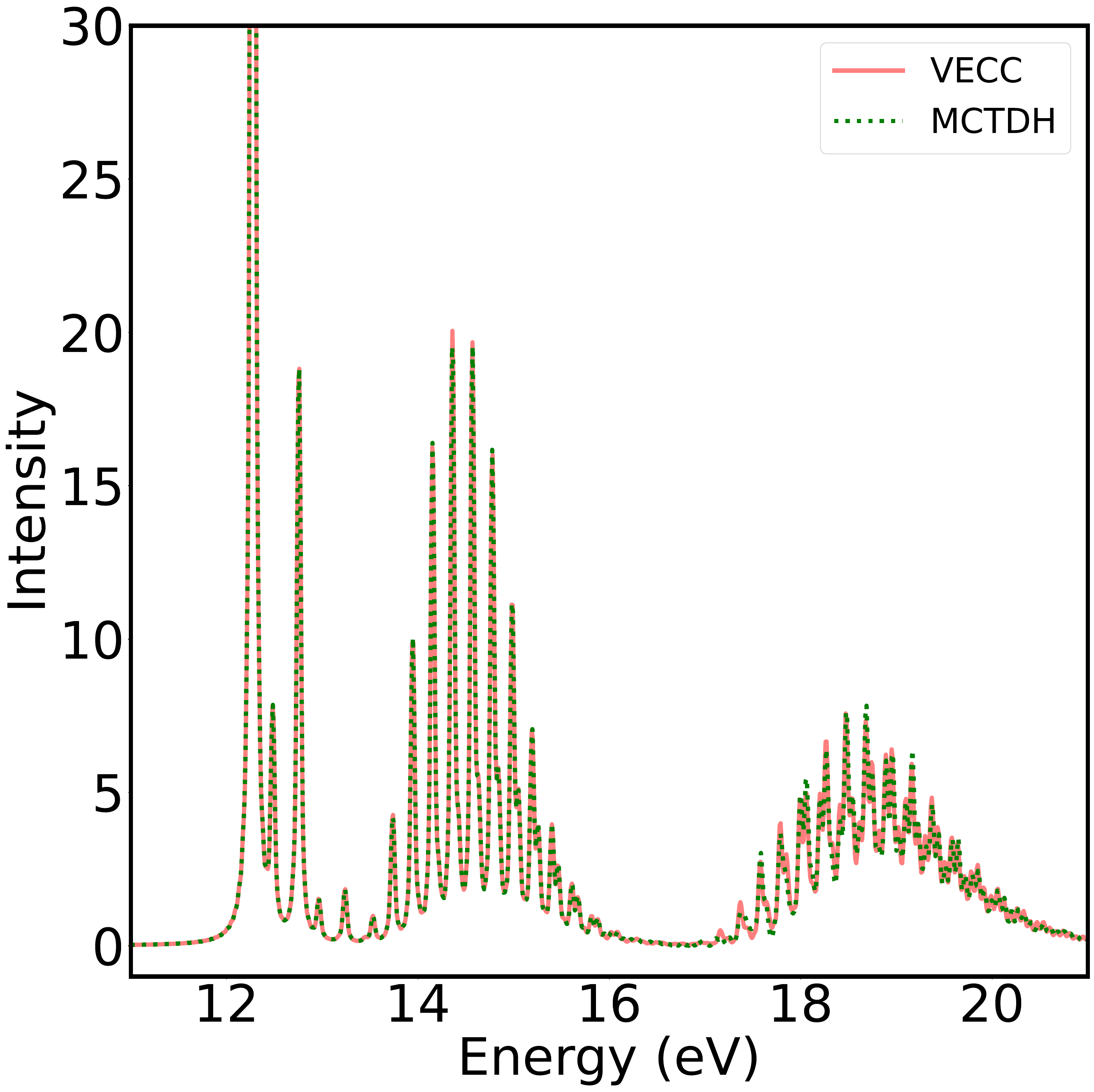}%
    }\hfill
    \subfloat[\label{subfig:ammonia_spectra}%
        Photo-electron spectra of ammonia after 100 fs propagation.%
    ]{%
        \includegraphics[width=0.8\columnwidth]{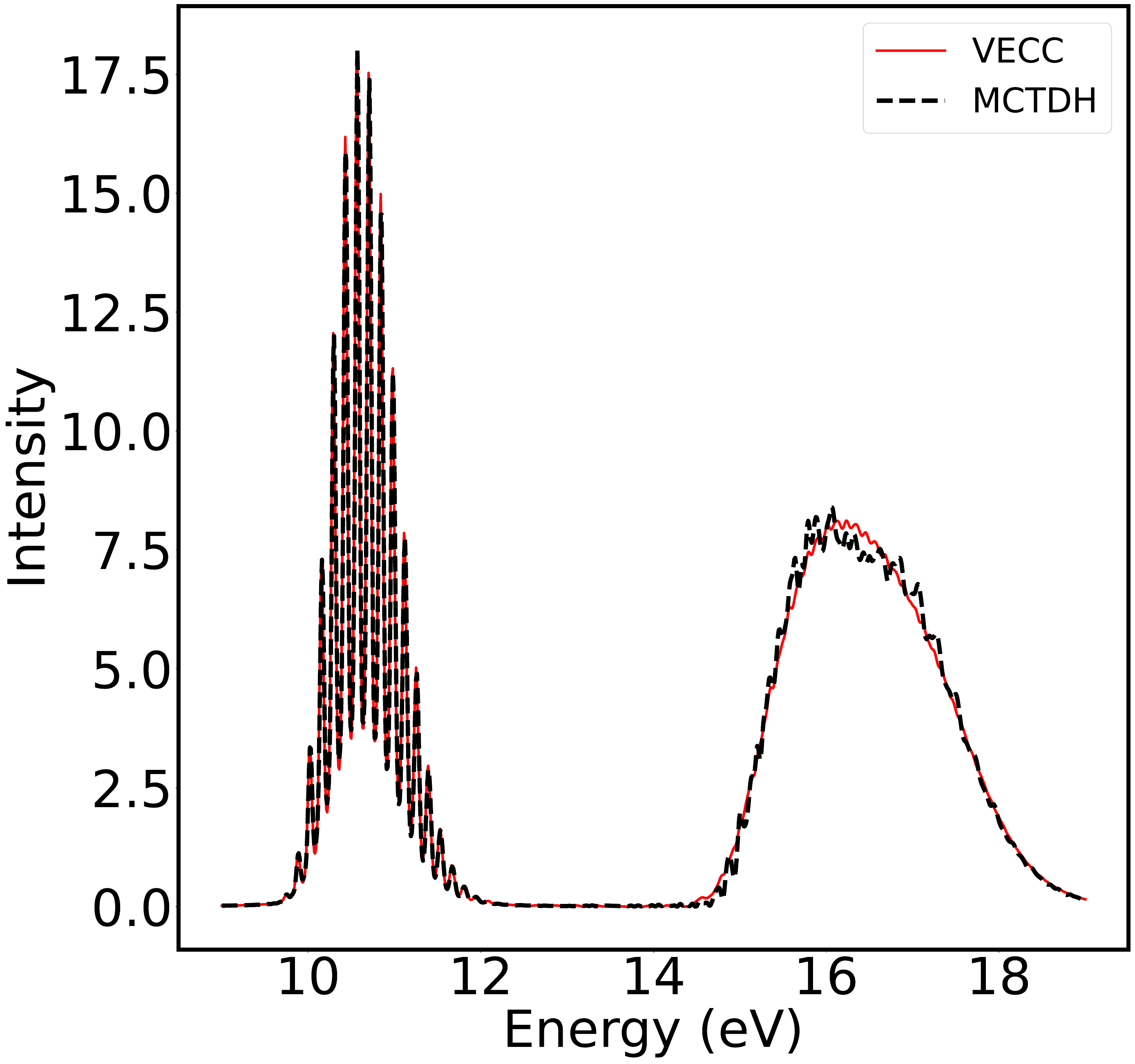}
    }\hfill
    \caption{%
        \label{fig:small_vibronic_models}
        Simulated absorption spectra of water and ammonia molecular compounds using VECC and MCTDH.
        Both are generated through a Fourier transform of their time-autocorrelation functions using~\texttt{autospec84}.
    }
\end{figure}

To further examine the performance of the VECC method, we consider vibronic models for two small molecular compounds.
As shown in \Fref{fig:small_vibronic_models}, we simulate the spectra for water and ammonia molecules and benchmark with the MCTDH method.

\Fref{subfig:water_spectra} shows the simulated spectra of the water molecule, which is a typical weak-coupling system.
The spectra has three bands; each band roughly matching the energy level of one excited electronic state.
The bands are broadened by excitations of vibrational levels.
In general the spectra match very well, although there are tiny discrepancies on a few peak intensities, we regard them as insignificant.~\footnote{%
Since these two methods are fundamentally different, we do not anticipate that they should produce the exact same result.
In particular MCTDH results are subject to truncations of a finite basis set.
}

\Fref{subfig:ammonia_spectra} shows the simulated spectra of the ammonia molecule, which is a typical strong-coupling system.
The spectra of the ammonia molecule, has two bands.
The higher energy band closely matches the energy level of the two degenerate Jahn-Teller states, and the lower energy band corresponds to the remaining non-degenerate state.
For the lower energy band, VECC almost exactly matches MCTDH.
In the higher energy band, VECC produces similar peak shape and width compared to MCTDH, but produces slightly different fine-structures, that may be hard to converge.

In general, we show that the VECC method can produce accurate vibrationally-resolved electronic spectra.
The VECC method has some difficulties predicting fine-structures consistent with MCTDH for vibronic models with strong coupling.
However, the accuracy is acceptable and consistent with experimental spectra, yielding semi-quantitative results.
Moreover, the VECC method has a significant advantage in terms of computational runtime compared to the MCTDH method, as well as its ease of use.

\subsubsection{\label{sec:Three_One_Three} Spectra for large molecular compounds}
The benchmarks for spectra of small molecular compounds demonstrate that the VECC method is capable of accurately simulating vibronic spectra.
To evaluate the efficiency of the VECC method, we benchmark absorption spectra for larger molecular compounds.

\begin{figure}[!h]\centering%
    \subfloat[\label{subfig:pyrazine_a}%
        Benchmark of absorption spectras generated by VECC against MCTDH, after 300 fs propagation%
    ]{
        \includegraphics[width=0.8\columnwidth]{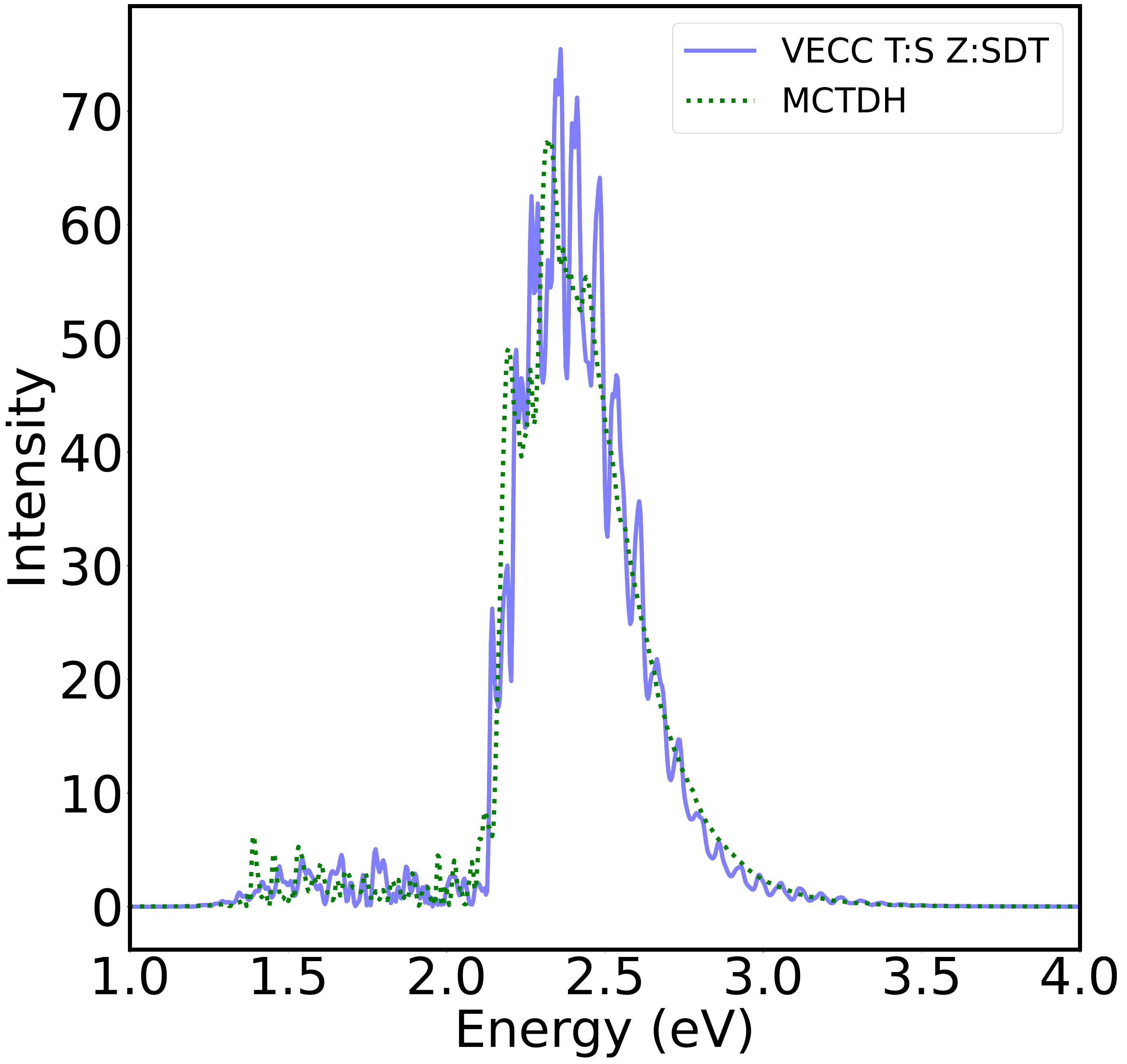}%
    }\hfill
    \subfloat[\label{subfig:pyrazine_b}%
        Comparison of absorption spectra simulated by VECC, at different CC truncation levels, after 300 fs propagation.%
    ]{%
        \includegraphics[width=0.8\columnwidth]{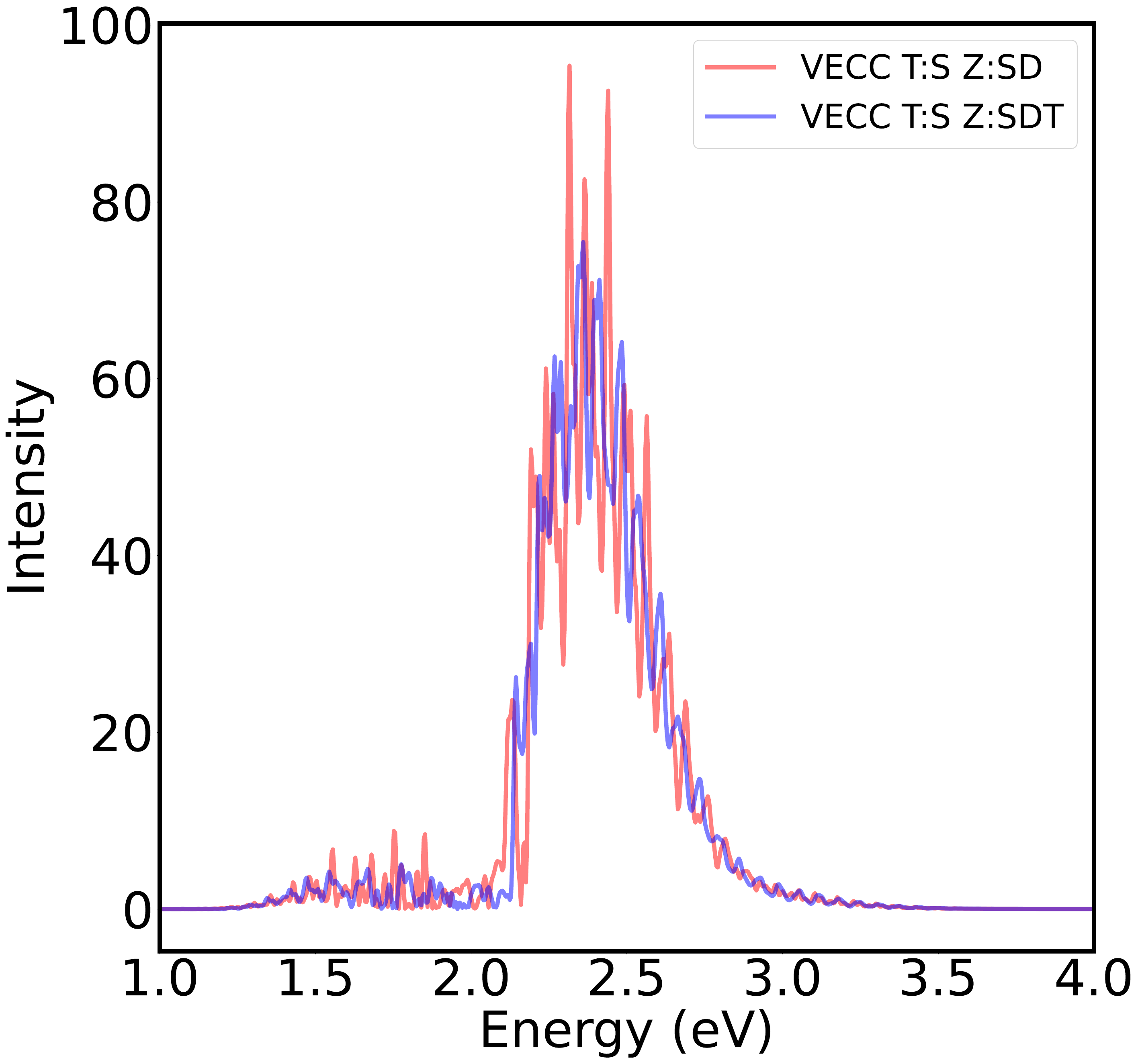}
    }\hfill
    \caption{%
        \label{fig:pyrazine_spectra}
        We simulate the absorption spectra of pyrazine (2 states, 24 modes) using the VECC and MCTDH methods.
        The model and Fourier transform parameters were obtained from the reference by Raab et al.~\cite{raab1999molecular}
    }
\end{figure}

The absorption spectra of pyrazine, shown in~\Fref{fig:pyrazine_spectra} was simulated using VECC and MCTDH.
The spectra has two bands: the lower energy band ranges from~\qtyrange{1.4}{2.0}{eV} and the higher energy band ranges from~\qtyrange{2.0}{3.0}{eV}.
The VECC spectra roughly matches the MCTDH spectra in both bands.
Compared to MCTDH, the VECC method produces sharper peaks and richer fine-structures in the higher energy band.
The cause of the discrepancies between the two methods can be complicated.
Note that, as shown in the reference by Raab et al, the MCTDH spectra fails to match the experimental spectra %
 without introducing empirical broadening of the peaks in the Fourier transform.~\cite{raab1999molecular}

In \Fref{subfig:pyrazine_b} we can see that as truncation order of the $Z$ amplitudes increase, the spectra converges towards the MCTDH spectra, especially with regards to the peak intensities.
The spectra results are close to each other in general.
We therefore conclude that we have sufficient convergence in our VECC calculation, in terms of the truncation level of the CC amplitudes.

VECC shows tremendous advantage in terms of computation runtime compared to MCTDH, between two to three orders of magnitude faster.
It takes $>1$ day (using an Intel Xeon E5-2667) to simulate the spectra of the pyrazine vibronic model with MCTDH.~\footnote{%
using the same input file settings as in the reference by Raab et al.~\cite{raab1999molecular}%
}
The VECC method, takes $\sim \SI{1}{minute}$ with doubles (SD) truncation of the $\hat{Z}$ amplitude and $\sim \SI{2.5}{minutes}$ with triples (SDT) truncation of the $\hat{Z}$ amplitude (using an Apple M2).

\begin{figure}\centering%
    \includegraphics[width=0.8\columnwidth]{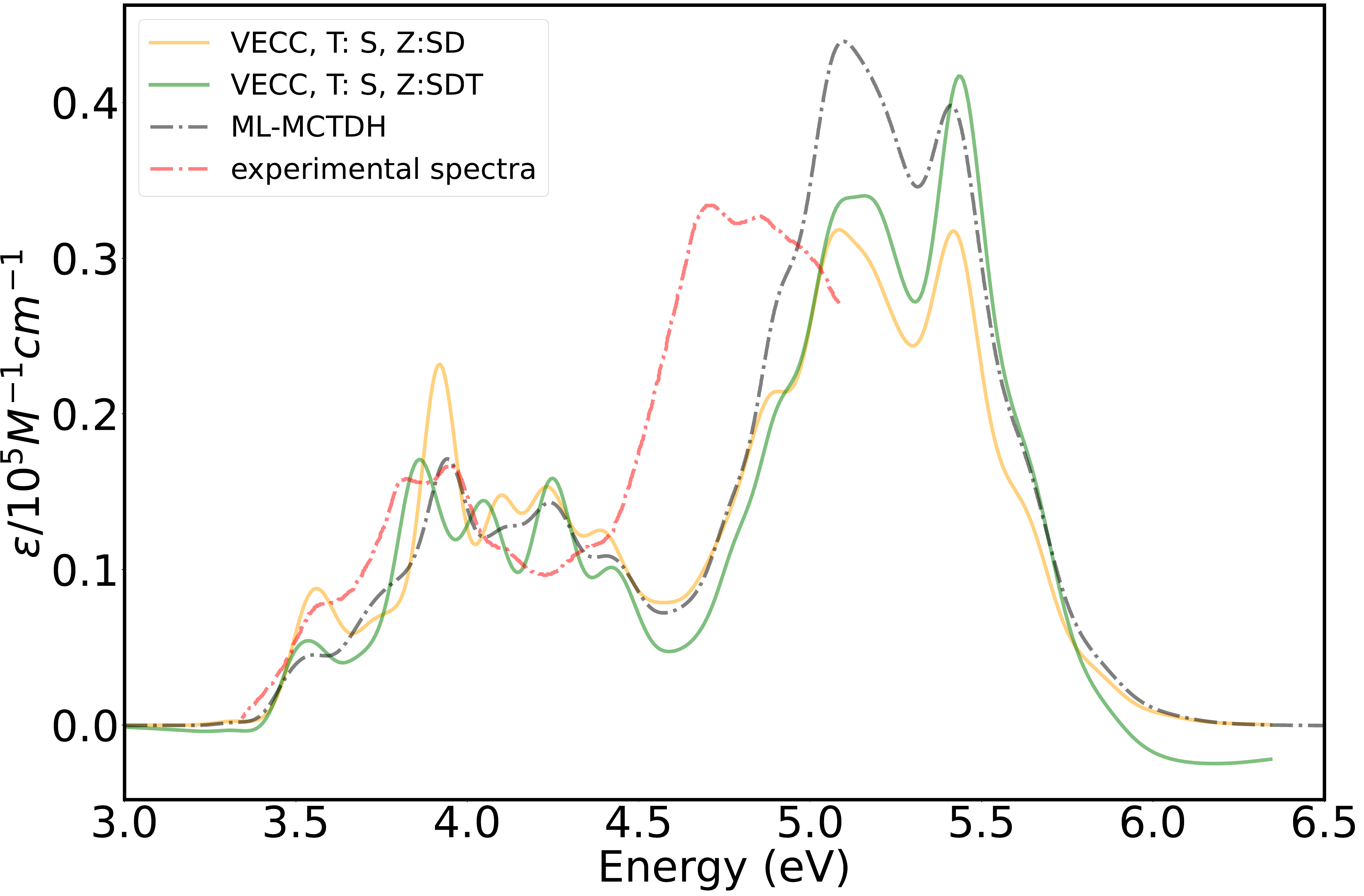}
    \caption{%
        \label{fig:hexa_spectra}%
        Simulated and experimental absorption spectra of hexahelicene (14 states, 63 modes).
        Comparing the VECC method against the ML-MCTDH method used in Aranda et al,~\cite{aranda2021vibronic} after 300 fs propagation.
    }
\end{figure}

The other large vibronic model we investigated was hexahelicene.
\Fref{fig:hexa_spectra} compares the absorption spectra produced from the VECC method,  the reference ML-MCTDH spectra and the experimental spectra as reported in \cite{aranda2021vibronic}.
In the spectra of hexahelicene, there are mainly two bands: the lower energy band ranges from~\qtyrange{3.5}{5.0}{eV} and the higher energy band ranges from~\qtyrange{5.0}{6.0}{eV}.
In the simulated spectra the vertical energies, zero point energy, and intensities are shifted according to the reference.

In the lower energy band, the VECC peak positions and intensities roughly match those of ML-MCTDH.
Moreover, as shown in the reference~\cite{aranda2021vibronic}, the Frank-Condon Hertzberg-Teller model with vertical gradient (FCHT\(\mid\)VG) method fails at this region where its prediction of the peak positions and intensities has qualitative different from the experimental spectra.
The correct prediction of peak positions and intensities illustrates that the VECC method manages to successfully incorporate the effect of vibronic coupling.

In the higher energy band, the peak positions from the VECC method are consistent with that of the ML-MCTDH method.
However, the peak intensities from the VECC calculation are slightly lower.
Interestingly, the ML-MCTDH method seems to overestimate the peak intensities in the higher energy band, and the VECC result is in fact closer to the experimental spectra.

The SDT and SD truncation results of the VECC method only show slight differences.
In comparison to the SD result, the peak intensities of the SDT result seem to converge towards the ML-MCTDH method.
The results indicate that the calculation has reached sufficient convergence even at SD truncation of the CC amplitudes.
We think it is a desirable feature of the VECC approach that both SD and SDT results can be obtained with reasonable computational costs, such that one can assess convergence.

The benchmark studies of absorption spectra for these two compounds verify the robustness of the VECC method for simulating spectra of large molecular compounds.

Compared to the ML-MCTDH method, which requires complicated settings of basis function and multi-layer tree structures, using the VECC method is much simpler, requiring only two inputs: vibronic model parameters and a choice of $\hat{T}$ and $\hat{Z}$ truncation levels.
With the attractive features of improved scaling and user friendly inputs, the VECC method is well positioned to be widely applied to simulations of photo-electron and UV-VIS absorption spectra.


\subsection{\label{sec:Three_Two} Diabatic state population}

Next we turn to the calculation of diabatic state populations.
We implemented the formulation described in~\Sref{sec:Two_Three_Three} for calculating diabatic state populations using VECC.
The diabatic state population is evaluated at each numerical integration step in an ``on-the-fly'' fashion.
Using the ammonia model from~\Sref{sec:Three_One_One} we compared VECC and MCTDH.

\begin{figure}\centering%
    \subfloat[\label{subfig:a}%
        Diabatic state population of ammonia calculated using the VECC method%
    ]{
        \includegraphics[width=0.8\columnwidth]{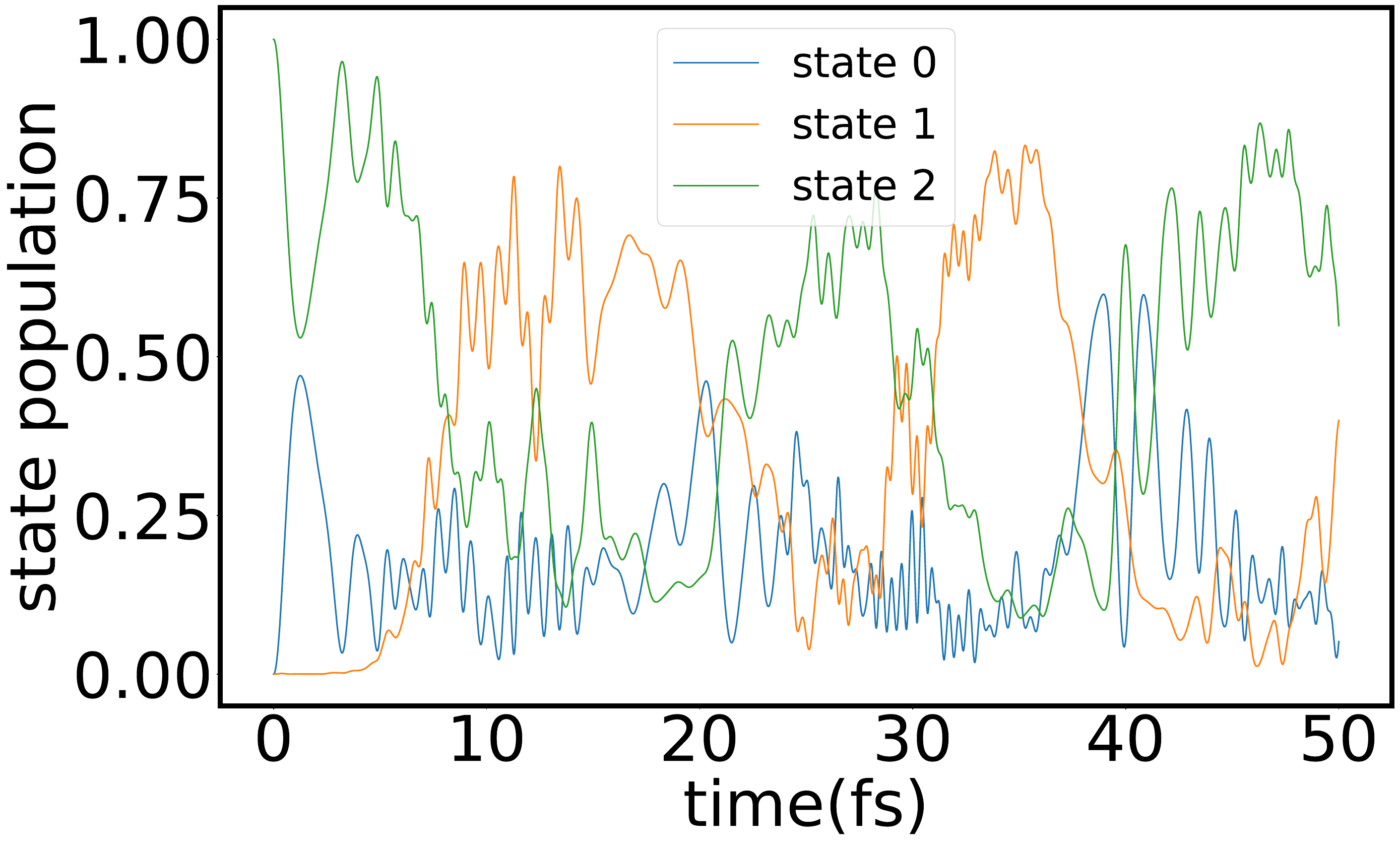}%
    }\hfill
    \subfloat[\label{subfig:b}%
        Diabatic state population of ammonia calculated using the MCTDH method%
    ]{
        \includegraphics[width=0.8\columnwidth]{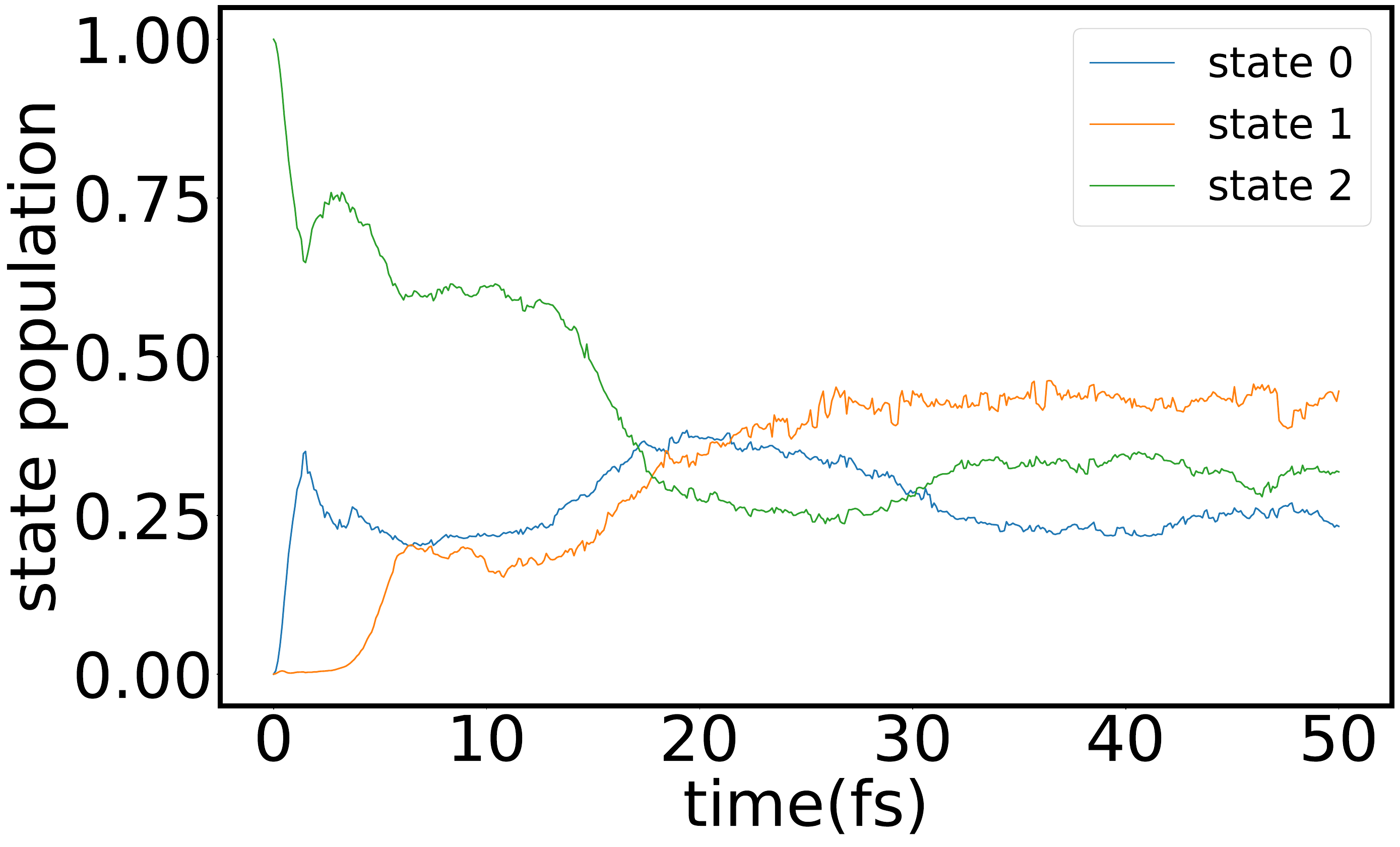}%
    }\hfill
    \caption{%
        \label{fig:ammonia_pop_2}%
        Comparison of the diabatic state populations of the ammonia model.
        Surprisingly, the VECC populations deviate significantly from those produced by MCTDH, even though their respective absorption spectras have almost exact agreement.
        In general, the VECC method produces state populations that have longer period of quantum coherence and have stronger Rabi oscillations than the MCTDH method.
    }
\end{figure}

\Fref{fig:ammonia_pop_2} shows the evolution of the diabatic state populations for ammonia, initialized at one of the two Jahn-Teller states.
Unfortunately, the state population produced using VECC fails to match that of MCTDH, even though the VECC method previously produced accurate absorption spectra, as seen in \Fref{subfig:ammonia_spectra}.
%
In general the VECC method produces state populations with longer periods of quantum coherence, as well as stronger Rabi oscillations than the MCTDH method.

The cause of this discrepancy is likely associated with the Ehrenfest parameterization described in \Sref{sec:Two_Two_Two}.
In this parameterization, we chose $\hat{T}$ to be independent of electronic states and weight-averaged over all electronic states.
As shown in \Eref{eqn:not_deco_ref} and \Eref{eqn:not_deco_state}, each diabatic state is determined by a linear cluster expansion based on a common reference state.

\begin{subequations}
    \begin{equation}\label{eqn:not_deco_ref}
        \ket{\text{ref}} = e^{\hat{T}(\tau)} \lsum{b}\ket{b, 0}
    \end{equation}
    \begin{equation}\label{eqn:not_deco_state}
        \ket{D_{b}} = \hat{Z}_{b}(\tau) \ket{\text{ref}}
    \end{equation}
\end{subequations}
%
For example, if we choose to truncate the $\hat{Z}$ amplitude at doubles (for each vibration mode) the different diabatic states could differ by, at most, two-fold excitations (in terms of the vibrational degrees of freedom).
This has the side effect of preserving the large overlap between the different diabatic states throughout the time evolution.
Thus, the VECC populations oscillate more strongly than the MCTDH populations.

In general, we anticipate that the current formulation of VECC method will produce more accurate results for state averaged time-dependent properties, while struggling to be as accurate for state specific time-dependent properties.

\section{\label{sec:Four} Concluding remarks}

\subsection{In Summary}
In this work, our goal was to develop a novel computational method that can efficiently simulate non-adiabatic dynamics based on a diabatic vibronic Hamiltonian.
Non-adiabatic dynamics have wide applications in simulations of photo-electron/UV-VIS spectra and diabatic state populations.

To achieve this goal we developed the VECC method, which can be regarded as a special form of the TNOE approach proposed by M. Nooijen and S. Bao.~\cite{nooijen2021normal}
A key benefit of the VECC method is that it allows us to solve the quantum dynamics problem without the need for introducing basis sets.
In the VECC method, we apply a second-quantized CC ansatz to parameterize the time-dependent wave-function.
To solve the quantum dynamics problem, we substitute the CC ansatz into the TDSE to form CC EOM.\@
The computational cost of evaluating these CC EOM scales classically as the size of the system.
In this way, the VECC method dramatically reduces the computational cost compared to conventional basis set based methods that have exponential scaling over the size of the system.

From our benchmark studies we verified that the VECC method is capable of simulating photo-electron and UV/VIS absorption spectra efficiently and accurately. It is likely that the accuracy of the parameterized (gas-phase) vibronic model is the primary candidate for discrepancies with experiment.
Thus, the VECC method in conjunction with vibronic models is well positioned to be widely applied to simulations of experimental spectra, which can facilitate the characterization of molecular compounds and understanding of their optical properties.

It is clear that the current implementation of VECC has some limitations. The electronic transition moments are assumed to be constant.
In our experience such an approach is well-justified for vibronic models as the diabatic states do not change much with geometry.
Additionally, we cannot presently simulate hot bands, but using the mixed CC/CI approach this should not be too complicated to include.
Importantly, the VECC approach has difficulties predicting accurate diabatic state populations for vibronic models with strong coupling.
One obvious obstacle is that VECC diabatic state populations experience stronger quantum coherence and for longer durations, when compared to MCTDH.
We attribute the primary cause of this loss in accuracy to our use of an Ehrenfest parameterization in \Sref{sec:Two_Two_Two} which assumes a single common important geometry (or moving Gaussian wave packet) for all electronic states.
Another limitation of the VECC method described in this work, is that we restrict our diabatic potential to a form of polynomial expansion that is truncated at quadratic terms.
This limits the application of the VECC method in circumstances which requires higher-order expansions of the PES and different forms of the potential (e.g.\ bond breaking).

Overall the VECC method performs quite impressively when simulating absorption spectra and presents significant improvements in computational runtime when compared to MCTDH. It is also convenient to use in contrast with ML-MCTDH which requires a lot of delicate input from users of the program.
Although with a few limitations, the VECC method proposed in this paper is one of the first attempts to apply the CC method to the multi-surface excited state dynamics problem.
This work opens a new field and has great potential for more future exploration.

\subsection{Final Remarks}
In this section we offer some insight regarding choices made during the development of the VECC theory,
as well as discuss connections with established methods from the literature.

\par
The most straightforward exponential parameterization for wave-functions associated with vibronic models would include a single operator $\hat{T}= \sum_{a,b} \hat{T}^a_b$, where each electronic component $\hat{T}^a_b$ has some normal mode component.
However, the matrix nature of such operators ($\hat{T}$) is problematic when evaluating the TDSE.
Computing $\frac{d}{d \tau} e^{\hat{T}}$ is complicated, as we cannot use the standard chain rule, and the corresponding similarity-transformed Hamiltonian ($\emT \hat{H} \eT$) is not a connected operator.
While we explored some of these possibilities in early work (unpublished), we abandoned them in favor of exponential operators that have no electronic dependence (or at most diagonal dependence).
As can be seen in \Eref{eqn:sim_tran_H_single_surface} and \Eref{eqn:sim_tran_H_multi_surface} their resulting similarity-transformed Hamiltonians are connected operators.

\par
For single surface Hamiltonians it suffices to use only creation operators in $\hat{T}$, and one can achieve exact results with the single-exponential ansatz (for up to quadratic Hamiltonians and second order terms in $\hat{T}$).
It is somewhat surprising that one can use non-unitary parameterizations to achieve exact results.
We found that the complex ``constant'' term ($t_0$)
is necessary to preserve the normalization and energy conservation properties. The value of this parameter is simply calculated from the propagation of the equation of motion.
However, as mentioned previously, extending such an exponential operator to multiple surfaces introduces many complications.
Therefore we explored other parameterizations and showed that a mixed-CC/CI parameterization ($e^{\hat{T}} \hat{Z}$) is sufficiently accurate in the single-surface case, and convenient and robust for the multi-surface case.

\par
The multi surface vibronic problem is quite sensitive to divergences while propagating the amplitude equations, which came as a surprise to us.
The robust mixed-CC/CI parameterization (see~Eqs.~(\ref{eqn:vib_ansatz}-\ref{eqn:vib_ansatz_p2})) was necessary to overcome these divergences.
However, this is a redundant parameterization where we have the freedom to choose the form of $\hat{T}$ or $\hat{Z}$.
For the mixed-CC/CI ansatz we chose to truncate $\hat{T}$ at singles (\Eref{eqn:vib_ansatz_p1}).
In principle, we could choose to truncate $\hat{T}$ at singles and doubles.
However, truncating at singles simplifies the calculation of properties for multiple surfaces (See~\Eref{eqn:BCH_exp_eT_dagger} in \Sref{sec:Two_Three_Two}).
We do not anticipate this choice to significantly affect results, because when $\hat{T}$ is truncated at singles we achieve sufficient accuracy for the single surface case, as shown in~\Frefpl{fig:three}{fig:four}.
We also impose a state averaged equation for $\hat{T}$, and this essentially describes translations of the coordinate system.
In the spirit of Ehrenfest dynamics, the state-averaged equation results in the same coordinates being used for each state.
If one takes the view that $\hat{T}$ causes a translation, it is natural to project against states that are also translated and this leads to modified projection manifold, or equivalently the doubly transformed Hamiltonian $\hat{G}=\eTd \emT \hat{H} \eT e^{-\hat{T}^{\dagger}}$.
However, it appears that a drawback of this Ehrenfest parameterization is that it causes longer periods of quantum coherence and
stronger Rabi oscillations when calculating diabatic state populations.

We think the mixed CC/CI parameterization can be further understood by outlining the similarity with a Gaussian wave packet approach. If we consider the VECC wave-function
\begin{equation}
 \ket{\Psi(t)} =\sum_x e^{T_1}\hat{Z}_x \ket{0,x}
\end{equation}
It could alternatively be represented in real space as
\begin{eqnarray}
 \ket{\Psi(q,t)} =\sum_x \big[ z_x^0 +
 \sum_i z_x^i (q_i-Q_i) + \nonumber \\
 \qquad \sum_{i,j} z_x^{ij}(q_i-Q_i)(q_j-Q_j) + \cdots \big] e^{- \frac{1}{2} \sum_i (q_i-Q_i)^2} \nonumber \\
\end{eqnarray}
The central coordinate $Q(t)$ of the Gaussian is generated by our $e^{T_1}$ and is the same for all electronic states. This Gaussian function is multiplied by a general state-specific polynomial, which would be quadratic for the SD model and cubic for the SDT model. Associating a swarm of cartesian polynomials with a moving Gaussian is an appealing picture, reminicent of atomic orbital basis sets used in electronic structure theory, but now in all vibrational dimensions. This conceptually convenient representation also naturally leads to the modified time-evolving projection manifold we use. The derivation of equations would appear to be more convenient using our second-quantized approach, avoiding the use of integrals.

This model could be improved by using state-specific centers $Q_x$. Likewise the current VECC approach might be improved by using state-specific $T_1$ amplitudes, e.g.
\begin{equation}
 \ket{\Psi(\tau)} =\sum_x e^{\hat{T}_{1,x}(\tau)}\hat{Z}_x(\tau) \ket{0,x}
\end{equation}
 The difficulties with such a method would be developing equations for the amplitudes, in particular when the electronic state has little population. These problems have been solved in Ab initio multiple spawning and Gaussian wave-packet approaches, ~\cite{ben2000ab, hudock2007ab, williams2021unmasking, mori2012role,curchod2016communication,burghardt2003multiconfigurational, richings2015quantum, christopoulou2021improved} . This can provide inspiration for generalizations of our current Ehrenfest based VECC approach.

%
%
%
%
%
%
%
%
%
%
%
%
%
%
%
\section*{Acknowledgements}
We thank Prof Fabrizio Santoro's group who provided the original data for their ML-MCTDH calculation, as well as the vibronic model parameters of the hexahelicene compound that we benchmarked against in \Sref{sec:Three_One}.
In particular, Daniel Aranda generously answered detailed questions about the diabatization procedure and computational costs.

We also thank Julia Endicott who constructed the vibronic models of ammonia and water.

We acknowledge the support of the Natural Sciences and Engineering Research Council of Canada (NSERC).

\section*{Author declarations}

\subsection*{Conflict of Interest}
The authors have no conflicts to disclose.

\subsection*{Author Contributions}
\textbf{Songhao Bao}: Conceptualization (equal), Investigation (equal), Methodology (equal), Software (supporting), Validation (equal), Writing - original draft (lead), Writing - review \& editing (equal)
\textbf{Neil Raymond}: Investigation (equal), Software (lead), Validation (equal), Writing - original draft (supporting), Writing - review \& editing (equal)
\textbf{Marcel Nooijen}: Conceptualization (lead),  Funding acquisition (lead), Methodology (lead), Supervision (lead), Writing - original draft (supporting), Writing - review \& editing (equal)

\section*{Data declarations}
The data that support the findings of this study are available from the corresponding author upon reasonable request.

\bibliographystyle{jcp}  
\bibliography{paper}


\end{document}